\documentclass[longauth]{aaEC}

\usepackage{graphicx}
\usepackage{natbib}
\usepackage{scalerel}
\setcitestyle{notesep={; }}
\usepackage[table]{xcolor}
\newenvironment{redblock}
  {\begingroup\color{black}}
  {\endgroup}
\usepackage{txfonts}
\usepackage[pdfencoding=auto,psdextra]{hyperref}
\hypersetup{
    colorlinks=true,
    linkcolor=blue,
    filecolor=magenta,      
    urlcolor=blue,
    citecolor=blue
}
\makeatletter
\renewcommand*\aa@pageof{, page \thepage{} of \pageref*{LastPage}}
\makeatother

\usepackage[utf8]{inputenc}

\usepackage[switch, modulo]{lineno}

\usepackage{euclid}

\begin{document}
%
%

\title{\Euclid: Improving redshift distribution reconstruction using a deep-to-wide transfer function\thanks{This paper is published on behalf of the Euclid Consortium.}}    

   
\newcommand{\orcid}[1]{} 
\author{Y.~Kang\orcid{0009-0000-8588-7250}\thanks{\email{yuzheng.kang@unige.ch}}\inst{\ref{aff1}}
\and S.~Paltani\orcid{0000-0002-8108-9179}\inst{\ref{aff1}}
\and W.~G.~Hartley\inst{\ref{aff1}}
\and M.~Bolzonella\orcid{0000-0003-3278-4607}\inst{\ref{aff2}}
\and A.~H.~Wright\orcid{0000-0001-7363-7932}\inst{\ref{aff3}}
\and F.~Dubath\orcid{0000-0002-6533-2810}\inst{\ref{aff1}}
\and F.~J.~Castander\orcid{0000-0001-7316-4573}\inst{\ref{aff4},\ref{aff5}}
\and D.~C.~Masters\orcid{0000-0001-5382-6138}\inst{\ref{aff6}}
\and W.~d'Assignies~\orcid{0000-0002-9719-1717}\inst{\ref{aff7}}
\and H.~Hildebrandt\orcid{0000-0002-9814-3338}\inst{\ref{aff3}}
\and O.~Ilbert\orcid{0000-0002-7303-4397}\inst{\ref{aff8}}
\and M.~Manera\orcid{0000-0003-4962-8934}\inst{\ref{aff9},\ref{aff7}}
\and W.~Roster\orcid{0000-0002-9149-6528}\inst{\ref{aff10}}
\and S.~A.~Stanford\orcid{0000-0003-0122-0841}\inst{\ref{aff11}}
\and N.~Aghanim\orcid{0000-0002-6688-8992}\inst{\ref{aff12}}
\and B.~Altieri\orcid{0000-0003-3936-0284}\inst{\ref{aff13}}
\and S.~Andreon\orcid{0000-0002-2041-8784}\inst{\ref{aff14}}
\and N.~Auricchio\orcid{0000-0003-4444-8651}\inst{\ref{aff2}}
\and H.~Aussel\orcid{0000-0002-1371-5705}\inst{\ref{aff15}}
\and C.~Baccigalupi\orcid{0000-0002-8211-1630}\inst{\ref{aff16},\ref{aff17},\ref{aff18},\ref{aff19}}
\and M.~Baldi\orcid{0000-0003-4145-1943}\inst{\ref{aff20},\ref{aff2},\ref{aff21}}
\and S.~Bardelli\orcid{0000-0002-8900-0298}\inst{\ref{aff2}}
\and P.~Battaglia\orcid{0000-0002-7337-5909}\inst{\ref{aff2}}
\and A.~Biviano\orcid{0000-0002-0857-0732}\inst{\ref{aff17},\ref{aff16}}
\and E.~Branchini\orcid{0000-0002-0808-6908}\inst{\ref{aff22},\ref{aff23},\ref{aff14}}
\and M.~Brescia\orcid{0000-0001-9506-5680}\inst{\ref{aff24},\ref{aff25}}
\and J.~Brinchmann\orcid{0000-0003-4359-8797}\inst{\ref{aff26},\ref{aff27},\ref{aff28}}
\and S.~Camera\orcid{0000-0003-3399-3574}\inst{\ref{aff29},\ref{aff30},\ref{aff31}}
\and G.~Ca\~nas-Herrera\orcid{0000-0003-2796-2149}\inst{\ref{aff32},\ref{aff33}}
\and V.~Capobianco\orcid{0000-0002-3309-7692}\inst{\ref{aff31}}
\and C.~Carbone\orcid{0000-0003-0125-3563}\inst{\ref{aff34}}
\and V.~F.~Cardone\inst{\ref{aff35},\ref{aff36}}
\and J.~Carretero\orcid{0000-0002-3130-0204}\inst{\ref{aff37},\ref{aff38}}
\and S.~Casas\orcid{0000-0002-4751-5138}\inst{\ref{aff39},\ref{aff40}}
\and M.~Castellano\orcid{0000-0001-9875-8263}\inst{\ref{aff35}}
\and G.~Castignani\orcid{0000-0001-6831-0687}\inst{\ref{aff2}}
\and S.~Cavuoti\orcid{0000-0002-3787-4196}\inst{\ref{aff25},\ref{aff41}}
\and K.~C.~Chambers\orcid{0000-0001-6965-7789}\inst{\ref{aff42}}
\and A.~Cimatti\inst{\ref{aff43}}
\and C.~Colodro-Conde\inst{\ref{aff44}}
\and G.~Congedo\orcid{0000-0003-2508-0046}\inst{\ref{aff32}}
\and L.~Conversi\orcid{0000-0002-6710-8476}\inst{\ref{aff45},\ref{aff13}}
\and Y.~Copin\orcid{0000-0002-5317-7518}\inst{\ref{aff46}}
\and A.~Costille\inst{\ref{aff8}}
\and F.~Courbin\orcid{0000-0003-0758-6510}\inst{\ref{aff47},\ref{aff48},\ref{aff49}}
\and H.~M.~Courtois\orcid{0000-0003-0509-1776}\inst{\ref{aff50}}
\and M.~Cropper\orcid{0000-0003-4571-9468}\inst{\ref{aff51}}
\and H.~Degaudenzi\orcid{0000-0002-5887-6799}\inst{\ref{aff1}}
\and G.~De~Lucia\orcid{0000-0002-6220-9104}\inst{\ref{aff17}}
\and H.~Dole\orcid{0000-0002-9767-3839}\inst{\ref{aff12}}
\and C.~A.~J.~Duncan\orcid{0009-0003-3573-0791}\inst{\ref{aff32}}
\and X.~Dupac\inst{\ref{aff13}}
\and S.~Dusini\orcid{0000-0002-1128-0664}\inst{\ref{aff52}}
\and A.~Ealet\orcid{0000-0003-3070-014X}\inst{\ref{aff46}}
\and S.~Escoffier\orcid{0000-0002-2847-7498}\inst{\ref{aff53}}
\and M.~Farina\orcid{0000-0002-3089-7846}\inst{\ref{aff54}}
\and R.~Farinelli\inst{\ref{aff2}}
\and S.~Farrens\orcid{0000-0002-9594-9387}\inst{\ref{aff15}}
\and F.~Faustini\orcid{0000-0001-6274-5145}\inst{\ref{aff35},\ref{aff55}}
\and S.~Ferriol\inst{\ref{aff46}}
\and F.~Finelli\orcid{0000-0002-6694-3269}\inst{\ref{aff2},\ref{aff56}}
\and N.~Fourmanoit\orcid{0009-0005-6816-6925}\inst{\ref{aff53}}
\and M.~Frailis\orcid{0000-0002-7400-2135}\inst{\ref{aff17}}
\and E.~Franceschi\orcid{0000-0002-0585-6591}\inst{\ref{aff2}}
\and M.~Fumana\orcid{0000-0001-6787-5950}\inst{\ref{aff34}}
\and S.~Galeotta\orcid{0000-0002-3748-5115}\inst{\ref{aff17}}
\and K.~George\orcid{0000-0002-1734-8455}\inst{\ref{aff57}}
\and B.~Gillis\orcid{0000-0002-4478-1270}\inst{\ref{aff32}}
\and C.~Giocoli\orcid{0000-0002-9590-7961}\inst{\ref{aff2},\ref{aff21}}
\and J.~Gracia-Carpio\inst{\ref{aff10}}
\and A.~Grazian\orcid{0000-0002-5688-0663}\inst{\ref{aff58}}
\and F.~Grupp\inst{\ref{aff10},\ref{aff59}}
\and S.~V.~H.~Haugan\orcid{0000-0001-9648-7260}\inst{\ref{aff60}}
\and H.~Hoekstra\orcid{0000-0002-0641-3231}\inst{\ref{aff33}}
\and W.~Holmes\inst{\ref{aff61}}
\and F.~Hormuth\inst{\ref{aff62}}
\and A.~Hornstrup\orcid{0000-0002-3363-0936}\inst{\ref{aff63},\ref{aff64}}
\and P.~Hudelot\inst{\ref{aff65}}
\and K.~Jahnke\orcid{0000-0003-3804-2137}\inst{\ref{aff66}}
\and M.~Jhabvala\inst{\ref{aff67}}
\and B.~Joachimi\orcid{0000-0001-7494-1303}\inst{\ref{aff68}}
\and E.~Keih\"anen\orcid{0000-0003-1804-7715}\inst{\ref{aff69}}
\and S.~Kermiche\orcid{0000-0002-0302-5735}\inst{\ref{aff53}}
\and A.~Kiessling\orcid{0000-0002-2590-1273}\inst{\ref{aff61}}
\and B.~Kubik\orcid{0009-0006-5823-4880}\inst{\ref{aff46}}
\and M.~K\"ummel\orcid{0000-0003-2791-2117}\inst{\ref{aff59}}
\and M.~Kunz\orcid{0000-0002-3052-7394}\inst{\ref{aff70}}
\and H.~Kurki-Suonio\orcid{0000-0002-4618-3063}\inst{\ref{aff71},\ref{aff72}}
\and R.~Laureijs\inst{\ref{aff73}}
\and A.~M.~C.~Le~Brun\orcid{0000-0002-0936-4594}\inst{\ref{aff74}}
\and S.~Ligori\orcid{0000-0003-4172-4606}\inst{\ref{aff31}}
\and P.~B.~Lilje\orcid{0000-0003-4324-7794}\inst{\ref{aff60}}
\and V.~Lindholm\orcid{0000-0003-2317-5471}\inst{\ref{aff71},\ref{aff72}}
\and I.~Lloro\orcid{0000-0001-5966-1434}\inst{\ref{aff75}}
\and G.~Mainetti\orcid{0000-0003-2384-2377}\inst{\ref{aff76}}
\and D.~Maino\inst{\ref{aff77},\ref{aff34},\ref{aff78}}
\and E.~Maiorano\orcid{0000-0003-2593-4355}\inst{\ref{aff2}}
\and O.~Mansutti\orcid{0000-0001-5758-4658}\inst{\ref{aff17}}
\and S.~Marcin\inst{\ref{aff79}}
\and O.~Marggraf\orcid{0000-0001-7242-3852}\inst{\ref{aff80}}
\and M.~Martinelli\orcid{0000-0002-6943-7732}\inst{\ref{aff35},\ref{aff36}}
\and N.~Martinet\orcid{0000-0003-2786-7790}\inst{\ref{aff8}}
\and F.~Marulli\orcid{0000-0002-8850-0303}\inst{\ref{aff81},\ref{aff2},\ref{aff21}}
\and R.~J.~Massey\orcid{0000-0002-6085-3780}\inst{\ref{aff82}}
\and E.~Medinaceli\orcid{0000-0002-4040-7783}\inst{\ref{aff2}}
\and S.~Mei\orcid{0000-0002-2849-559X}\inst{\ref{aff83},\ref{aff84}}
\and Y.~Mellier\thanks{Deceased}\inst{\ref{aff85},\ref{aff65}}
\and M.~Meneghetti\orcid{0000-0003-1225-7084}\inst{\ref{aff2},\ref{aff21}}
\and E.~Merlin\orcid{0000-0001-6870-8900}\inst{\ref{aff35}}
\and G.~Meylan\inst{\ref{aff86}}
\and A.~Mora\orcid{0000-0002-1922-8529}\inst{\ref{aff87}}
\and M.~Moresco\orcid{0000-0002-7616-7136}\inst{\ref{aff81},\ref{aff2}}
\and L.~Moscardini\orcid{0000-0002-3473-6716}\inst{\ref{aff81},\ref{aff2},\ref{aff21}}
\and R.~Nakajima\orcid{0009-0009-1213-7040}\inst{\ref{aff80}}
\and C.~Neissner\orcid{0000-0001-8524-4968}\inst{\ref{aff7},\ref{aff38}}
\and S.-M.~Niemi\orcid{0009-0005-0247-0086}\inst{\ref{aff88}}
\and C.~Padilla\orcid{0000-0001-7951-0166}\inst{\ref{aff7}}
\and F.~Pasian\orcid{0000-0002-4869-3227}\inst{\ref{aff17}}
\and K.~Pedersen\inst{\ref{aff89}}
\and V.~Pettorino\orcid{0000-0002-4203-9320}\inst{\ref{aff88}}
\and S.~Pires\orcid{0000-0002-0249-2104}\inst{\ref{aff15}}
\and G.~Polenta\orcid{0000-0003-4067-9196}\inst{\ref{aff55}}
\and M.~Poncet\inst{\ref{aff90}}
\and L.~A.~Popa\inst{\ref{aff91}}
\and L.~Pozzetti\orcid{0000-0001-7085-0412}\inst{\ref{aff2}}
\and F.~Raison\orcid{0000-0002-7819-6918}\inst{\ref{aff10}}
\and A.~Renzi\orcid{0000-0001-9856-1970}\inst{\ref{aff92},\ref{aff52}}
\and J.~Rhodes\orcid{0000-0002-4485-8549}\inst{\ref{aff61}}
\and G.~Riccio\inst{\ref{aff25}}
\and E.~Romelli\orcid{0000-0003-3069-9222}\inst{\ref{aff17}}
\and M.~Roncarelli\orcid{0000-0001-9587-7822}\inst{\ref{aff2}}
\and R.~Saglia\orcid{0000-0003-0378-7032}\inst{\ref{aff59},\ref{aff10}}
\and Z.~Sakr\orcid{0000-0002-4823-3757}\inst{\ref{aff93},\ref{aff94},\ref{aff95}}
\and A.~G.~S\'anchez\orcid{0000-0003-1198-831X}\inst{\ref{aff10}}
\and D.~Sapone\orcid{0000-0001-7089-4503}\inst{\ref{aff96}}
\and B.~Sartoris\orcid{0000-0003-1337-5269}\inst{\ref{aff59},\ref{aff17}}
\and P.~Schneider\orcid{0000-0001-8561-2679}\inst{\ref{aff80}}
\and T.~Schrabback\orcid{0000-0002-6987-7834}\inst{\ref{aff97}}
\and A.~Secroun\orcid{0000-0003-0505-3710}\inst{\ref{aff53}}
\and G.~Seidel\orcid{0000-0003-2907-353X}\inst{\ref{aff66}}
\and S.~Serrano\orcid{0000-0002-0211-2861}\inst{\ref{aff5},\ref{aff98},\ref{aff4}}
\and P.~Simon\inst{\ref{aff80}}
\and C.~Sirignano\orcid{0000-0002-0995-7146}\inst{\ref{aff92},\ref{aff52}}
\and G.~Sirri\orcid{0000-0003-2626-2853}\inst{\ref{aff21}}
\and L.~Stanco\orcid{0000-0002-9706-5104}\inst{\ref{aff52}}
\and J.~Steinwagner\orcid{0000-0001-7443-1047}\inst{\ref{aff10}}
\and P.~Tallada-Cresp\'{i}\orcid{0000-0002-1336-8328}\inst{\ref{aff37},\ref{aff38}}
\and A.~N.~Taylor\inst{\ref{aff32}}
\and I.~Tereno\orcid{0000-0002-4537-6218}\inst{\ref{aff99},\ref{aff100}}
\and N.~Tessore\orcid{0000-0002-9696-7931}\inst{\ref{aff51}}
\and S.~Toft\orcid{0000-0003-3631-7176}\inst{\ref{aff101},\ref{aff102}}
\and R.~Toledo-Moreo\orcid{0000-0002-2997-4859}\inst{\ref{aff103}}
\and F.~Torradeflot\orcid{0000-0003-1160-1517}\inst{\ref{aff38},\ref{aff37}}
\and I.~Tutusaus\orcid{0000-0002-3199-0399}\inst{\ref{aff4},\ref{aff5},\ref{aff94}}
\and J.~Valiviita\orcid{0000-0001-6225-3693}\inst{\ref{aff71},\ref{aff72}}
\and T.~Vassallo\orcid{0000-0001-6512-6358}\inst{\ref{aff17}}
\and A.~Veropalumbo\orcid{0000-0003-2387-1194}\inst{\ref{aff14},\ref{aff23},\ref{aff22}}
\and Y.~Wang\orcid{0000-0002-4749-2984}\inst{\ref{aff104}}
\and J.~Weller\orcid{0000-0002-8282-2010}\inst{\ref{aff59},\ref{aff10}}
\and G.~Zamorani\orcid{0000-0002-2318-301X}\inst{\ref{aff2}}
\and F.~M.~Zerbi\inst{\ref{aff14}}
\and I.~A.~Zinchenko\orcid{0000-0002-2944-2449}\inst{\ref{aff105}}
\and E.~Zucca\orcid{0000-0002-5845-8132}\inst{\ref{aff2}}
\and J.~Garc\'ia-Bellido\orcid{0000-0002-9370-8360}\inst{\ref{aff106}}
\and J.~Mart\'{i}n-Fleitas\orcid{0000-0002-8594-569X}\inst{\ref{aff107}}
\and V.~Scottez\orcid{0009-0008-3864-940X}\inst{\ref{aff85},\ref{aff108}}
\and M.~Viel\orcid{0000-0002-2642-5707}\inst{\ref{aff16},\ref{aff17},\ref{aff19},\ref{aff18},\ref{aff109}}
\and R.~Teyssier\orcid{0000-0001-7689-0933}\inst{\ref{aff110}}}
										   
\institute{Department of Astronomy, University of Geneva, ch. d'Ecogia 16, 1290 Versoix, Switzerland\label{aff1}
\and
INAF-Osservatorio di Astrofisica e Scienza dello Spazio di Bologna, Via Piero Gobetti 93/3, 40129 Bologna, Italy\label{aff2}
\and
Ruhr University Bochum, Faculty of Physics and Astronomy, Astronomical Institute (AIRUB), German Centre for Cosmological Lensing (GCCL), 44780 Bochum, Germany\label{aff3}
\and
Institute of Space Sciences (ICE, CSIC), Campus UAB, Carrer de Can Magrans, s/n, 08193 Barcelona, Spain\label{aff4}
\and
Institut d'Estudis Espacials de Catalunya (IEEC),  Edifici RDIT, Campus UPC, 08860 Castelldefels, Barcelona, Spain\label{aff5}
\and
Infrared Processing and Analysis Center, California Institute of Technology, Pasadena, CA 91125, USA\label{aff6}
\and
Institut de F\'{i}sica d'Altes Energies (IFAE), The Barcelona Institute of Science and Technology, Campus UAB, 08193 Bellaterra (Barcelona), Spain\label{aff7}
\and
Aix-Marseille Universit\'e, CNRS, CNES, LAM, Marseille, France\label{aff8}
\and
Serra H\'unter Fellow, Departament de F\'{\i}sica, Universitat Aut\`onoma de Barcelona, E-08193 Bellaterra, Spain\label{aff9}
\and
Max Planck Institute for Extraterrestrial Physics, Giessenbachstr. 1, 85748 Garching, Germany\label{aff10}
\and
Department of Physics and Astronomy, University of California, Davis, CA 95616, USA\label{aff11}
\and
Universit\'e Paris-Saclay, CNRS, Institut d'astrophysique spatiale, 91405, Orsay, France\label{aff12}
\and
ESAC/ESA, Camino Bajo del Castillo, s/n., Urb. Villafranca del Castillo, 28692 Villanueva de la Ca\~nada, Madrid, Spain\label{aff13}
\and
INAF-Osservatorio Astronomico di Brera, Via Brera 28, 20122 Milano, Italy\label{aff14}
\and
Universit\'e Paris-Saclay, Universit\'e Paris Cit\'e, CEA, CNRS, AIM, 91191, Gif-sur-Yvette, France\label{aff15}
\and
IFPU, Institute for Fundamental Physics of the Universe, via Beirut 2, 34151 Trieste, Italy\label{aff16}
\and
INAF-Osservatorio Astronomico di Trieste, Via G. B. Tiepolo 11, 34143 Trieste, Italy\label{aff17}
\and
INFN, Sezione di Trieste, Via Valerio 2, 34127 Trieste TS, Italy\label{aff18}
\and
SISSA, International School for Advanced Studies, Via Bonomea 265, 34136 Trieste TS, Italy\label{aff19}
\and
Dipartimento di Fisica e Astronomia, Universit\`a di Bologna, Via Gobetti 93/2, 40129 Bologna, Italy\label{aff20}
\and
INFN-Sezione di Bologna, Viale Berti Pichat 6/2, 40127 Bologna, Italy\label{aff21}
\and
Dipartimento di Fisica, Universit\`a di Genova, Via Dodecaneso 33, 16146, Genova, Italy\label{aff22}
\and
INFN-Sezione di Genova, Via Dodecaneso 33, 16146, Genova, Italy\label{aff23}
\and
Department of Physics "E. Pancini", University Federico II, Via Cinthia 6, 80126, Napoli, Italy\label{aff24}
\and
INAF-Osservatorio Astronomico di Capodimonte, Via Moiariello 16, 80131 Napoli, Italy\label{aff25}
\and
Instituto de Astrof\'isica e Ci\^encias do Espa\c{c}o, Universidade do Porto, CAUP, Rua das Estrelas, PT4150-762 Porto, Portugal\label{aff26}
\and
Faculdade de Ci\^encias da Universidade do Porto, Rua do Campo de Alegre, 4150-007 Porto, Portugal\label{aff27}
\and
European Southern Observatory, Karl-Schwarzschild-Str.~2, 85748 Garching, Germany\label{aff28}
\and
Dipartimento di Fisica, Universit\`a degli Studi di Torino, Via P. Giuria 1, 10125 Torino, Italy\label{aff29}
\and
INFN-Sezione di Torino, Via P. Giuria 1, 10125 Torino, Italy\label{aff30}
\and
INAF-Osservatorio Astrofisico di Torino, Via Osservatorio 20, 10025 Pino Torinese (TO), Italy\label{aff31}
\and
Institute for Astronomy, University of Edinburgh, Royal Observatory, Blackford Hill, Edinburgh EH9 3HJ, UK\label{aff32}
\and
Leiden Observatory, Leiden University, Einsteinweg 55, 2333 CC Leiden, The Netherlands\label{aff33}
\and
INAF-IASF Milano, Via Alfonso Corti 12, 20133 Milano, Italy\label{aff34}
\and
INAF-Osservatorio Astronomico di Roma, Via Frascati 33, 00078 Monteporzio Catone, Italy\label{aff35}
\and
INFN-Sezione di Roma, Piazzale Aldo Moro, 2 - c/o Dipartimento di Fisica, Edificio G. Marconi, 00185 Roma, Italy\label{aff36}
\and
Centro de Investigaciones Energ\'eticas, Medioambientales y Tecnol\'ogicas (CIEMAT), Avenida Complutense 40, 28040 Madrid, Spain\label{aff37}
\and
Port d'Informaci\'{o} Cient\'{i}fica, Campus UAB, C. Albareda s/n, 08193 Bellaterra (Barcelona), Spain\label{aff38}
\and
Institute for Theoretical Particle Physics and Cosmology (TTK), RWTH Aachen University, 52056 Aachen, Germany\label{aff39}
\and
Deutsches Zentrum f\"ur Luft- und Raumfahrt e. V. (DLR), Linder H\"ohe, 51147 K\"oln, Germany\label{aff40}
\and
INFN section of Naples, Via Cinthia 6, 80126, Napoli, Italy\label{aff41}
\and
Institute for Astronomy, University of Hawaii, 2680 Woodlawn Drive, Honolulu, HI 96822, USA\label{aff42}
\and
Dipartimento di Fisica e Astronomia "Augusto Righi" - Alma Mater Studiorum Universit\`a di Bologna, Viale Berti Pichat 6/2, 40127 Bologna, Italy\label{aff43}
\and
Instituto de Astrof\'{\i}sica de Canarias, E-38205 La Laguna, Tenerife, Spain\label{aff44}
\and
European Space Agency/ESRIN, Largo Galileo Galilei 1, 00044 Frascati, Roma, Italy\label{aff45}
\and
Universit\'e Claude Bernard Lyon 1, CNRS/IN2P3, IP2I Lyon, UMR 5822, Villeurbanne, F-69100, France\label{aff46}
\and
Institut de Ci\`{e}ncies del Cosmos (ICCUB), Universitat de Barcelona (IEEC-UB), Mart\'{i} i Franqu\`{e}s 1, 08028 Barcelona, Spain\label{aff47}
\and
Instituci\'o Catalana de Recerca i Estudis Avan\c{c}ats (ICREA), Passeig de Llu\'{\i}s Companys 23, 08010 Barcelona, Spain\label{aff48}
\and
Institut de Ciencies de l'Espai (IEEC-CSIC), Campus UAB, Carrer de Can Magrans, s/n Cerdanyola del Vall\'es, 08193 Barcelona, Spain\label{aff49}
\and
UCB Lyon 1, CNRS/IN2P3, IUF, IP2I Lyon, 4 rue Enrico Fermi, 69622 Villeurbanne, France\label{aff50}
\and
Mullard Space Science Laboratory, University College London, Holmbury St Mary, Dorking, Surrey RH5 6NT, UK\label{aff51}
\and
INFN-Padova, Via Marzolo 8, 35131 Padova, Italy\label{aff52}
\and
Aix-Marseille Universit\'e, CNRS/IN2P3, CPPM, Marseille, France\label{aff53}
\and
INAF-Istituto di Astrofisica e Planetologia Spaziali, via del Fosso del Cavaliere, 100, 00100 Roma, Italy\label{aff54}
\and
Space Science Data Center, Italian Space Agency, via del Politecnico snc, 00133 Roma, Italy\label{aff55}
\and
INFN-Bologna, Via Irnerio 46, 40126 Bologna, Italy\label{aff56}
\and
University Observatory, LMU Faculty of Physics, Scheinerstr.~1, 81679 Munich, Germany\label{aff57}
\and
INAF-Osservatorio Astronomico di Padova, Via dell'Osservatorio 5, 35122 Padova, Italy\label{aff58}
\and
Universit\"ats-Sternwarte M\"unchen, Fakult\"at f\"ur Physik, Ludwig-Maximilians-Universit\"at M\"unchen, Scheinerstr.~1, 81679 M\"unchen, Germany\label{aff59}
\and
Institute of Theoretical Astrophysics, University of Oslo, P.O. Box 1029 Blindern, 0315 Oslo, Norway\label{aff60}
\and
Jet Propulsion Laboratory, California Institute of Technology, 4800 Oak Grove Drive, Pasadena, CA, 91109, USA\label{aff61}
\and
Felix Hormuth Engineering, Goethestr. 17, 69181 Leimen, Germany\label{aff62}
\and
Technical University of Denmark, Elektrovej 327, 2800 Kgs. Lyngby, Denmark\label{aff63}
\and
Cosmic Dawn Center (DAWN), Denmark\label{aff64}
\and
Institut d'Astrophysique de Paris, UMR 7095, CNRS, and Sorbonne Universit\'e, 98 bis boulevard Arago, 75014 Paris, France\label{aff65}
\and
Max-Planck-Institut f\"ur Astronomie, K\"onigstuhl 17, 69117 Heidelberg, Germany\label{aff66}
\and
NASA Goddard Space Flight Center, Greenbelt, MD 20771, USA\label{aff67}
\and
Department of Physics and Astronomy, University College London, Gower Street, London WC1E 6BT, UK\label{aff68}
\and
Department of Physics and Helsinki Institute of Physics, Gustaf H\"allstr\"omin katu 2, University of Helsinki, 00014 Helsinki, Finland\label{aff69}
\and
Universit\'e de Gen\`eve, D\'epartement de Physique Th\'eorique and Centre for Astroparticle Physics, 24 quai Ernest-Ansermet, CH-1211 Gen\`eve 4, Switzerland\label{aff70}
\and
Department of Physics, P.O. Box 64, University of Helsinki, 00014 Helsinki, Finland\label{aff71}
\and
Helsinki Institute of Physics, Gustaf H{\"a}llstr{\"o}min katu 2, University of Helsinki, 00014 Helsinki, Finland\label{aff72}
\and
Kapteyn Astronomical Institute, University of Groningen, PO Box 800, 9700 AV Groningen, The Netherlands\label{aff73}
\and
Laboratoire d'etude de l'Univers et des phenomenes eXtremes, Observatoire de Paris, Universit\'e PSL, Sorbonne Universit\'e, CNRS, 92190 Meudon, France\label{aff74}
\and
SKAO, Jodrell Bank, Lower Withington, Macclesfield SK11 9FT, UK\label{aff75}
\and
Centre de Calcul de l'IN2P3/CNRS, 21 avenue Pierre de Coubertin 69627 Villeurbanne Cedex, France\label{aff76}
\and
Dipartimento di Fisica "Aldo Pontremoli", Universit\`a degli Studi di Milano, Via Celoria 16, 20133 Milano, Italy\label{aff77}
\and
INFN-Sezione di Milano, Via Celoria 16, 20133 Milano, Italy\label{aff78}
\and
University of Applied Sciences and Arts of Northwestern Switzerland, School of Computer Science, 5210 Windisch, Switzerland\label{aff79}
\and
Universit\"at Bonn, Argelander-Institut f\"ur Astronomie, Auf dem H\"ugel 71, 53121 Bonn, Germany\label{aff80}
\and
Dipartimento di Fisica e Astronomia "Augusto Righi" - Alma Mater Studiorum Universit\`a di Bologna, via Piero Gobetti 93/2, 40129 Bologna, Italy\label{aff81}
\and
Department of Physics, Institute for Computational Cosmology, Durham University, South Road, Durham, DH1 3LE, UK\label{aff82}
\and
Universit\'e Paris Cit\'e, CNRS, Astroparticule et Cosmologie, 75013 Paris, France\label{aff83}
\and
CNRS-UCB International Research Laboratory, Centre Pierre Bin\'etruy, IRL2007, CPB-IN2P3, Berkeley, USA\label{aff84}
\and
Institut d'Astrophysique de Paris, 98bis Boulevard Arago, 75014, Paris, France\label{aff85}
\and
Institute of Physics, Laboratory of Astrophysics, Ecole Polytechnique F\'ed\'erale de Lausanne (EPFL), Observatoire de Sauverny, 1290 Versoix, Switzerland\label{aff86}
\and
Telespazio UK S.L. for European Space Agency (ESA), Camino bajo del Castillo, s/n, Urbanizacion Villafranca del Castillo, Villanueva de la Ca\~nada, 28692 Madrid, Spain\label{aff87}
\and
European Space Agency/ESTEC, Keplerlaan 1, 2201 AZ Noordwijk, The Netherlands\label{aff88}
\and
DARK, Niels Bohr Institute, University of Copenhagen, Jagtvej 155, 2200 Copenhagen, Denmark\label{aff89}
\and
Centre National d'Etudes Spatiales -- Centre spatial de Toulouse, 18 avenue Edouard Belin, 31401 Toulouse Cedex 9, France\label{aff90}
\and
Institute of Space Science, Str. Atomistilor, nr. 409 M\u{a}gurele, Ilfov, 077125, Romania\label{aff91}
\and
Dipartimento di Fisica e Astronomia "G. Galilei", Universit\`a di Padova, Via Marzolo 8, 35131 Padova, Italy\label{aff92}
\and
Institut f\"ur Theoretische Physik, University of Heidelberg, Philosophenweg 16, 69120 Heidelberg, Germany\label{aff93}
\and
Institut de Recherche en Astrophysique et Plan\'etologie (IRAP), Universit\'e de Toulouse, CNRS, UPS, CNES, 14 Av. Edouard Belin, 31400 Toulouse, France\label{aff94}
\and
Universit\'e St Joseph; Faculty of Sciences, Beirut, Lebanon\label{aff95}
\and
Departamento de F\'isica, FCFM, Universidad de Chile, Blanco Encalada 2008, Santiago, Chile\label{aff96}
\and
Universit\"at Innsbruck, Institut f\"ur Astro- und Teilchenphysik, Technikerstr. 25/8, 6020 Innsbruck, Austria\label{aff97}
\and
Satlantis, University Science Park, Sede Bld 48940, Leioa-Bilbao, Spain\label{aff98}
\and
Departamento de F\'isica, Faculdade de Ci\^encias, Universidade de Lisboa, Edif\'icio C8, Campo Grande, PT1749-016 Lisboa, Portugal\label{aff99}
\and
Instituto de Astrof\'isica e Ci\^encias do Espa\c{c}o, Faculdade de Ci\^encias, Universidade de Lisboa, Tapada da Ajuda, 1349-018 Lisboa, Portugal\label{aff100}
\and
Cosmic Dawn Center (DAWN)\label{aff101}
\and
Niels Bohr Institute, University of Copenhagen, Jagtvej 128, 2200 Copenhagen, Denmark\label{aff102}
\and
Universidad Polit\'ecnica de Cartagena, Departamento de Electr\'onica y Tecnolog\'ia de Computadoras,  Plaza del Hospital 1, 30202 Cartagena, Spain\label{aff103}
\and
Caltech/IPAC, 1200 E. California Blvd., Pasadena, CA 91125, USA\label{aff104}
\and
Astronomisches Rechen-Institut, Zentrum f\"ur Astronomie der Universit\"at Heidelberg, M\"onchhofstr. 12-14, 69120 Heidelberg, Germany\label{aff105}
\and
Instituto de F\'isica Te\'orica UAM-CSIC, Campus de Cantoblanco, 28049 Madrid, Spain\label{aff106}
\and
Aurora Technology for European Space Agency (ESA), Camino bajo del Castillo, s/n, Urbanizacion Villafranca del Castillo, Villanueva de la Ca\~nada, 28692 Madrid, Spain\label{aff107}
\and
ICL, Junia, Universit\'e Catholique de Lille, LITL, 59000 Lille, France\label{aff108}
\and
ICSC - Centro Nazionale di Ricerca in High Performance Computing, Big Data e Quantum Computing, Via Magnanelli 2, Bologna, Italy\label{aff109}
\and
Department of Astrophysical Sciences, Peyton Hall, Princeton University, Princeton, NJ 08544, USA\label{aff110}}

%
%
\abstract{The \Euclid\ mission of the European Space Agency seeks to understand the Universe’s expansion history and the nature of dark energy, through measurements of cosmic shear. This requires a very accurate estimate of the true redshift distribution of the galaxies, with the systematic error in the mean redshift satisfying $\sigma_{\langle z\rangle}<0.002(1+z)$ per tomographic bin. Achieving this accuracy relies on reference samples with spectroscopic redshifts, together with a procedure to match them to survey sources for which only photometric redshifts are available. One important source of systematic uncertainty is the mismatch in photometric properties between galaxies in the \Euclid survey and the reference objects. We develop a method to degrade the photometry of objects with deep photometry to match the properties of any shallower survey in the multi-band photometric space, preserving all the correlations between the fluxes and their uncertainties. We compare our transfer method with more demanding image-based methods, such as Balrog from the Dark Energy Survey Collaboration. According to our metrics, our method outperforms Balrog. We implement our method in the redshift distribution reconstruction, based on the self-organising map approach of \citet{Masters_2015}, and test it using a realistic sample from the \Euclid\ Flagship Mock Galaxy Simulation. We find that the key ingredient is to ensure that the reference objects are distributed in the colour space the same way as the wide-survey objects, which can be efficiently achieved with our transfer method. In our best implementation, the mean redshift biases are consistently reduced across the tomographic bins, bringing a significant fraction of them within the Euclid accuracy requirements in all tomographic bins. Equally importantly, the tests allow us to pinpoint which step in the calibration pipeline has the strongest impact on achieving the required accuracy. Our approach also reproduces the overall redshift distributions, which are crucial for applications such as angular clustering. The agreement between the reconstructed and true distributions demonstrates both the feasibility and robustness of the approach. This implementation is sufficient for Euclid Data Release 1 and provides a solid foundation for subsequent data releases.}
%
%
%
%
\titlerunning{Redshift calibration improvement using a deep-to-wide transfer function}
\authorrunning{Kang et al.}

\maketitle
%
%
%
%

\section{\label{sc:Intro}Introduction}
Many cosmological experiments have demonstrated that the Universe is undergoing accelerated expansion, with an equation of state of dark energy consistent with the cosmological constant (\citealp{planck_2020}, \citealp{2023PhRvD.107h3504A}, \citealp{Ghirardini2024}). The confirmation of any evidence for a deviation from this value, such as the recent studies from the Dark Energy Spectroscopic Instrument (DESI) Collaboration \citep{Adame_2025}, would have a significant impact on our understanding of the Universe. Measuring this equation of state and its evolution is therefore one of the major goals of modern cosmology. To achieve this goal and deepen our understanding of the dark Universe, modern surveys employ a variety of complementary probes, including weak gravitational lensing tomography, baryon acoustic oscillations, galaxy clustering, and galaxy clusters. Among these, weak lensing tomography \citep{Hu_1999}, which uses the shear of galaxy images induced by the passage of light through the cosmic web, has emerged as a powerful method to trace the evolution of the power spectrum of matter. However, it imposes some very stringent requirements on the measurement of the cosmic shear and on the accuracy of the redshift estimates \citep{Amara_2008}.

\Euclid\ is a medium-class mission led by the European Space Agency, specifically designed to advance this goal. It will survey approximately 14\,000\,deg$^2$ of the extragalactic sky, delivering high-resolution galaxy shapes in the optical \IE\ band with the VIS imaging instrument \citep{EuclidSkyVIS} and near-infrared spectroscopy and photometry in the \YE, \JE, and \HE\ bands with the NISP near-infrared instrument \citep{EuclidSkyNISP}. The Euclid Wide Survey \citep[EWS;][]{Scaramella-EP1} nominally reaches a depth of $\IE = 24.5$ at signal-to-noise ratio $\mathrm{S/N}=10$ for extended sources \citep{Scaramella-EP1, Q1-TP002}. These data are expected to constrain the quasi-linear approximation of the $w$ parameter of the equation of state $p=w(z)\,\rho$, with $w(z)=w_0 + w_\mathrm{a}\,z/(1+z)$, and $z$ the redshift, to a 1$\sigma$ precision of 0.02 for $w_0$ and 0.1 for $w_\mathrm{a}$. To meet these scientific objectives, the mission imposes stringent requirements on the knowledge of the redshift distribution $n(z)$ in at least 10 tomographic bins, with a systematic error on the mean redshift in each tomographic bin $\sigma_{\langle z \rangle} < 0.002(1+z)$ \citep{EuclidSkyOverview}. Given the vast sky coverage of the EWS and its depth, photometric redshifts are the only feasible approach for assigning redshifts to the billions of galaxies in the EWS. 

Photometric redshifts are derived by the \Euclid scientific data processing pipeline, which is designed to support both cosmology and non-cosmology science, such as galaxy evolution. This pipeline receives multi-band photometric catalogues produced upstream \citep{Q1-TP004} and provides for each object an estimate of the redshift, the source classification, the reconstructed spectral energy distribution (SED), and estimates of the source's physical parameters. It also constructs the $n(z)$ redshift distributions in each bin required for weak lensing, and, more generally, for 3\texttimes2pt cosmological inference. The Quick Release 1 version of the pipeline is described in \cite{Q1-TP005}, while significant changes will be implemented for Data Release 1, in particular to support cosmological science.

Photometric redshifts offer much higher completeness, especially for faint sources, than spectroscopic redshifts. However, they are inherently subject to biases arising from different sources. In particular, the SED templates, in the case of template-fitting algorithms, or the training sample, in the case of machine-learning algorithms, are never fully representative of the true population. These biases can propagate into cosmological analyses, where accurate redshift distributions are critical. As a result, meeting the stringent accuracy requirements using photometric redshifts alone remains a considerable challenge \citep{Bordoloi_2010}. This motivates the need for additional procedures that can mitigate the systematic offsets and improve the reliability of the reconstructed redshift distribution used in cosmological inference, a process often referred to as `calibration'.

Modern cosmological surveys have employed a range of methods for photometric redshift calibration. One common method is the direct calibration with spectroscopic redshifts originally proposed by \citet{Lima_2008}. Its basic principle is that the $n(z)$ of two galaxy samples should be identical if the two samples have the same magnitude and colour distributions. The method uses a spectroscopic sample where objects are weighted according to their prevalence in the second sample, whose $n(z)$ we want to measure. However, the matching of the samples in a high-dimensional colour space becomes computationally and conceptually more and more complex as the number of photometric bands increases.

To address this issue, \citet{Masters_2015} introduced a novel approach using self-organising maps \citep[SOM;][]{Kohonen_1990}. The SOM is an algorithm in the category of manifold learning, which provides a two-dimensional view of the complex, multi-dimensional manifold of galaxy colours, onto which spectroscopic and photometric samples can be projected. Within this framework, the redshift distribution in each cell is estimated using the number-weighted spectroscopic sample, which serves as a proxy for the true $n(z)$. This method improves upon the original direct calibration method by providing a visual and interpretable framework for the matching in colour space. Each cell is naturally occupied with galaxies with similar SEDs, and empty cells can be easily identified and either discarded or set up for follow-up observations. This was successfully demonstrated as part of the Complete Calibration of the Color-Redshift Relation (C3R2) programme \citep{Masters_2017}.

The SOM calibration implicitly assumes that the photometric depth, luminosity distribution, and data quality of the spectroscopic sample match those of the sample to be calibrated. The KiDS survey \citep{de_Jong_2013} benefits from extensive spectroscopic coverage overlapping directly with the survey. As a result, KiDS constrained redshift biases to $\sigma_{\langle z \rangle} \leq 0.006$ across all tomographic bins, supported by spectroscopic data covering 99\% of the full colour space of the photometric sample \citep{Wright_2020}. In other cases, such as the Dark Energy Survey (DES; \citealt{DES_2016}), the spectroscopic sample originates from a different field with deeper observations and possibly additional passbands. Thus, its photometric properties cannot match those of the Wide survey, in particular because of the larger S/N. The SOMPZ method \citep{Myles_2021} developed for the Dark Energy Survey addresses this  issue by constructing separate SOMs for the Wide and Deep samples, connecting them through a transfer function that predicts how a Deep-field source would appear when observed in the Wide survey. This information is then propagated to the wide-field SOM and used to calibrate the redshift distributions of wide-field galaxies. This approach enabled DES to achieve an uncertainty on the mean redshift in each tomographic bin $\sigma_{\langle z \rangle} \sim 0.01$  \citep{Myles_2021}. The transfer function from the Deep to the Wide SOMs is implemented using the Balrog method \citep{Everett_2022}, which involves injecting synthetic objects from the Deep sample into raw Wide imaging data that are then processed the same way as any other real source.

\begin{figure}[t]
    \centering
    \includegraphics[angle=0,width=1\hsize]{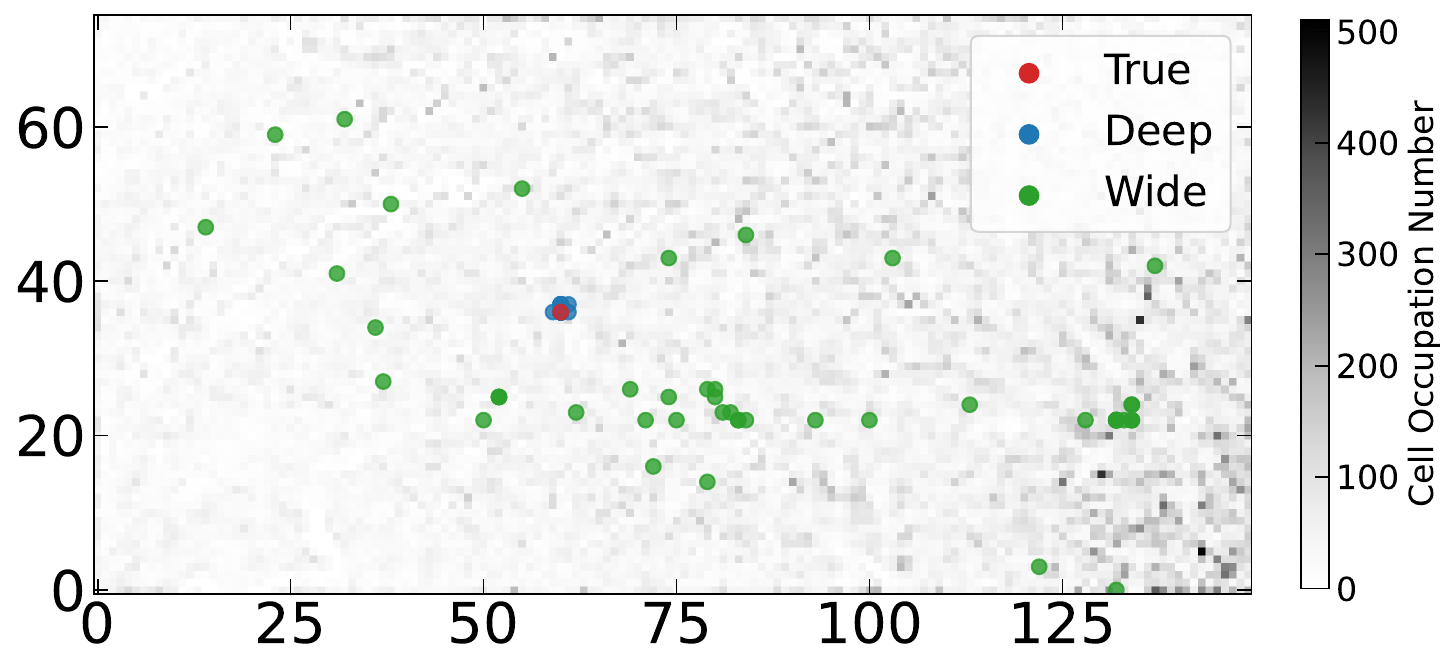}
    \caption{Projection of sources onto the self-organising map (SOM) trained using 8-band photometric data. Each SOM cell consists of objects with similar spectral energy distributions (SEDs). The background grayscale indicates the number of objects mapped to each cell. The red dot marks the true flux of a selected object projected onto the SOM, while the blue and green markers show 50 independent realisations of the same object with Deep and Wide photometric noise, respectively.
}
    \label{fig:deep_to_wide}
\end{figure}

In the framework of the \Euclid project, several studies on the redshift calibration are explored in parallel, including using the SOM \citep{Roster_2025} and clustering redshift \citep{dAssignies25}. \citet{Roster_2025} focuses on the issue of the definition of the tomographic bins. Here, we focus on the mismatch between the photometric properties of the sources with spectroscopic redshifts and those used to determine the shear. The EWS is indeed in a situation similar to DES, where the photometric redshift calibration pipeline relies on the existence of specific sky areas with deep multi-band photometry and numerous spectroscopic-redshift measurements obtained from complementary surveys, the Euclid Auxiliary Fields \citep[EAFs;][]{EuclidSkyOverview}, including AEGIS \citep{2007ApJ...660L...1D}, CDFS \citep{Giacconi_2001}, COSMOS \citep{Scoville_2007}, the \Euclid\ self-calibration field, ultra-deep field, GOODS-North \citep{2004ApJ...600L..93G}, SXDS \citep{Furusawa_2008}, and VVDS \citep{LeFevre_2005}. The EWS is approximately two magnitudes shallower than the EAFs. Thus, there are important differences in S/N. As a result, because of the different scatter in colours due to the uncertainties, identical sources observed in the EWS and the EAFs cover different regions of the SOM, as illustrated in Fig.~\ref{fig:deep_to_wide}. As in the case of DES, it is necessary to transform the Deep photometry to mimic wide-like observations before calibration. Implementing image-level simulations, as done in DES using Balrog, would, in principle, represent the most comprehensive approach to perform the transfer. However, such methods are computationally expensive and require substantial modifications to the very complex \Euclid pipeline. We propose here a catalogue-level transformation that offers a fast and effective solution to determine the transfer function without significant additional processing or substantial modifications to the pipeline. It is important to note that our goal is not merely to reproduce wide-like measurements starting from a deep survey. Instead, we aim at studying the impact of the transfer function on the $n(z)$ calibration.

This paper is structured as follows. In Sect.~\ref{sc:Transferring a deep data into wide-like}, we introduce the multi-passband transformation method and verify it using DES data. In Sect.~\ref{sc:nz_calibration}, we apply this method to test its impact on the calibration of the redshift distribution in the context of \Euclid. Section.~\ref{sc:discussion} presents a detailed discussion of the method, its performance, and broader applicability to cosmological analyses. Finally, we summarise our main conclusions in Sect.~\ref{sc:Conclusion}.

\section{\label{sc:Transferring a deep data into wide-like} Transferring a deep-sample object into a wide-like one}
As previously explained, because of the photometric uncertainties, a given source has a probability distribution of falling in a given SOM cell that depends on the depth of the observations, with deeper observations resulting in a more concentrated probability distribution around the true SOM cell, that is, the cell in which the source would fall if its true fluxes were known. Assuming the tomographic bins are constructed from a set .1of SOM cells, the same source observed with both Deep and Wide photometry may end up in different bins (see Fig.~\ref{fig:deep_to_wide}), inducing differences in the $n(z)$ distributions of the cells. Thus, when constructing the SOM with sources that have deep photometry, their fluxes need to be degraded so that they are scattered across the SOM in the same way as if the sources were observed at Wide depths. For this to happen, it is crucial that not only do the uncertainty distributions match, but also that all correlations between fluxes, uncertainties, and fluxes and uncertainties are preserved. We develop here an empirical method to accomplish this transformation.

\subsection{\label{sc:gaussian_projection}Single-passband transfer}
In many applications, a straightforward method is often employed to transform synthetic or deep data into wide-like observations using a fixed Gaussian error on the flux in each passband. We first add some realism by drawing the error from its distribution in the Wide sample. We call this method the single-passband transfer function (SPT). For a given object observed in the Deep sample with observed flux and uncertainty $(d_{X},\,\sigma_{X})$ in passband ${X}$, the likelihood $\mathcal{L}$ that this object is identical to a source in the Wide sample with observed flux and flux error $(w_X,\tau_X)$ is given by

\begin{equation}
    \label{eq:likelihood}
    \begin{aligned}
        \mathcal{L}
        &= \frac{1}{\sqrt{2\pi}\sqrt{\sigma^2_{X}+\tau^2_{X}}}\exp\left(-\frac{(d_{X}-w_{X})^2}{2(\sigma^2_{X}+\tau^2_{X})}\right)\,.
    \end{aligned}
\end{equation}

We compute the likelihood $\mathcal{L}$ for all objects in the Wide sample, and draw one wide-sample object $(w_X,\tau_X)$ at random following these likelihoods. The wide-like flux of the deep-sample source is given by $w^\prime_{X}\sim\mathcal{N}(d_{X}, \tau_{X}^2)$, where $\mathcal{N}(a,b)$ is the normal distribution with mean $a$ and variance $b$. We iterate on all passbands to degrade the deep-sample object into a wide-like one, each time potentially obtaining the wide-sample uncertainty from a different object. Applying this to all objects in the Deep sample, we obtain a degraded sample with one-dimensional error distributions matching those of the Wide survey, provided that the assumption that $\sigma_{X}$ is negligible compared to $\tau_{X}$ is true.

\subsection{Multi-passband transfer\label{sc:multiband}}
The SPT approach cannot capture the photometric correlations across the different quantities in the different bands. Poisson noise introduces a correlation between fluxes and flux errors, while flux errors across different passbands may be correlated due to the observing strategy or the local background. A realistic method to transfer a Deep object into a wide-like object needs to preserve these correlations.

For this purpose, we propose a multi-passband transfer (MPT) method\footnote{The code is available at \url{https://github.com/yuzheng-cosmos/multi-passband-transfer}.}. We consider a deep-sample object, with a set of fluxes and flux errors $(d_{{X},i},\,\sigma_{{X},i})$ in passband ${X}_i,\ i=1,\dots,N$. For each wide-sample object with flux and flux error $(w_{{X},i},\, \tau_{{X},i})$ in passband ${X}_i$, we compute the likelihood $\mathcal{L}$ that the object is identical to the deep-sample one

\begin{equation}
    \label{eq:likelihood_multi}
    \begin{aligned}
        \mathcal{L}
        &= \prod_{i=1}^N\frac{1}{\sqrt{2\pi}\sqrt{\sigma^2_{{X},i}+\tau^2_{{X},i}}}\exp\left(-\frac{(d_{{X},i}-w_{{X},i})^2}{2(\sigma^2_{{X},i}+\tau^2_{{X},i})}\right)\,.
    \end{aligned}
\end{equation}
   
We then draw a single wide-sample object at random following these likelihoods. The procedure can be repeated to produce as many realisations of the transfer as needed. The flux errors $\tau_{{X},i}$ from the selected neighbour(s) are then applied to the deep-sample object, and its new wide-like fluxes are obtained from the normal distributions

    \begin{equation}
        \label{eq:transfer}
        w^\prime_{{X},i} \sim \mathcal{N}(d_{{X},i}, \tau_{{X},i}^2)\,.
    \end{equation}
This method ensures that the applied flux errors in the multiple passbands contain all the correlations present in the Wide sample.
    
While one should in principle compute the likelihood in Eq.~(\ref{eq:likelihood_multi}) for all wide-sample objects, this is in general extremely inefficient, since most wide-sample sources have very small likelihoods to match the current deep-sample object. To speed up the processing, we restrict the computation of the likelihood, and hence the drawing of the flux errors, to wide-sample objects that are a priori likely to have large likelihoods. These objects are the nearest neighbours of the deep-sample source in flux space, neglecting flux uncertainties. They can be identified very efficiently using a k-d tree \citep{Bentley_1975,SciPy_2020}. The choice of the number of nearest neighbours for which the likelihoods are computed is driven by the computational efficiency on the one side and the accuracy of the transfer on the other side. We have found (see Sect.~\ref{sc:DES}) that using the 50 nearest neighbours is sufficient to obtain an accurate transfer.

    \begin{figure*}[tbp!]
        \centering
        \includegraphics[angle=0,width=1.0\hsize]{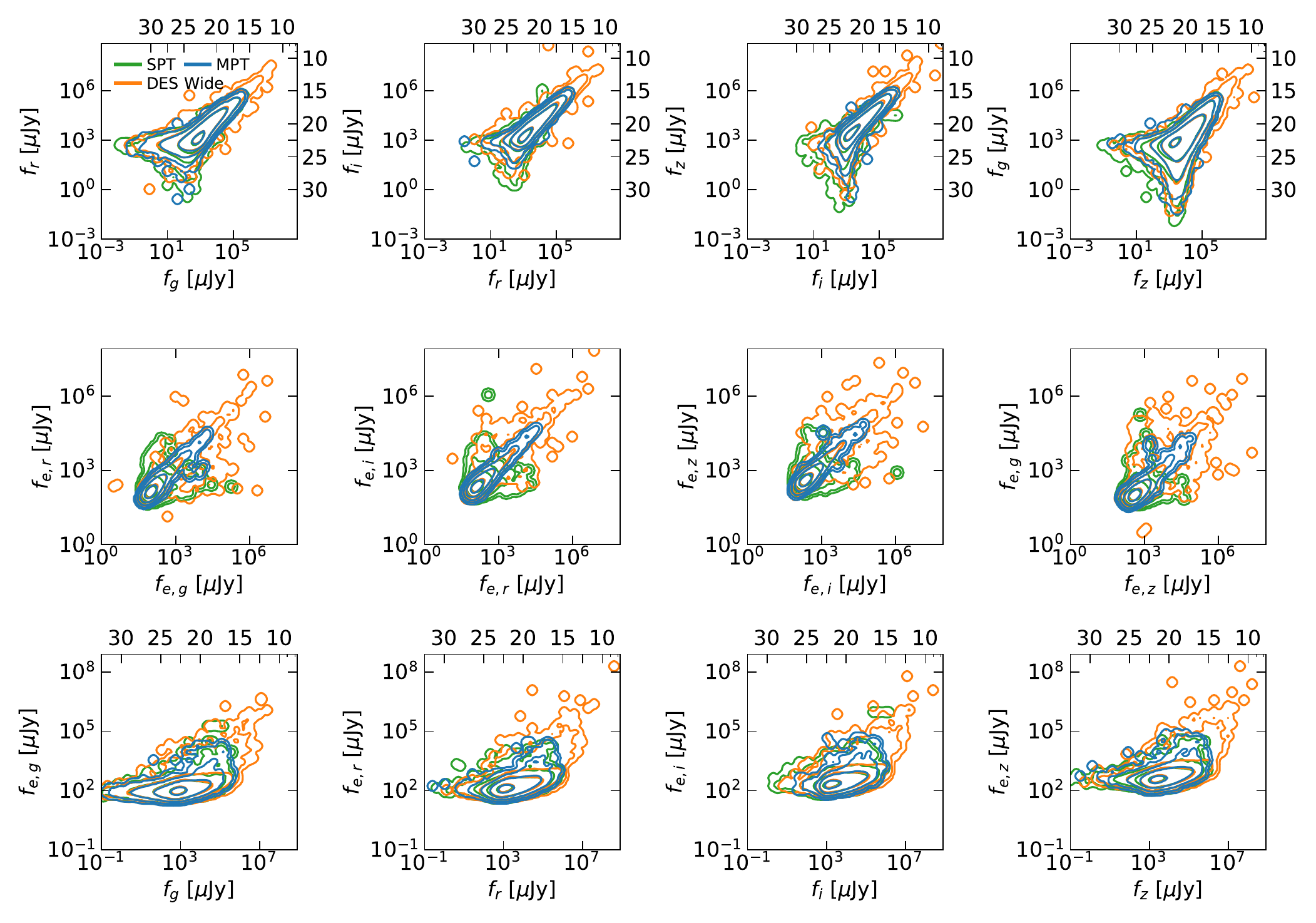}
        \caption{Comparison between the DES Wide data set and our transformations in flux and colour space. The figure shows the distribution of the DES Wide data set (orange), compared with the multi-passband transfer data set introduced in this paper (blue), and the wide-like data set generated using the single-passband transfer (green), where $f_X$ and $f_{e,X}$ indicate flux and flux error in passband $X$. The number on the top or right axis of the sub-figures indicates the corresponding magnitude.}
        \label{fig:compare_balrog}
    \end{figure*}
\begin{redblock}
\subsection{\label{sc:selection} Selection function}

Both the SPT and MPT methods convert all deep-sample objects into their wide-like counterparts. However, the deep-sample objects fainter than the wide-survey detection limit would not appear if they were observed at wide depth. To take this into account, after applying the deep-to-wide transformation, we introduce an empirical selection function derived from the wide-field data. 

We use the distribution of the wide-field flux in the detection band X, $\mathrm{d}N_\mathrm{X}/\mathrm{d}f_\mathrm{X}$, to construct an empirical selection function. We fit the high-flux part of $\mathrm{d}N_\mathrm{X}/\mathrm{d}f_\mathrm{X}$ with a power law with a fixed slope of $-2.5$. We then divide $\mathrm{d}N_\mathrm{X}/\mathrm{d}f_\mathrm{X}$ by this power law model to obtain an empirical selection function of the wide field, randomly accepting objects with a probability given by this function.
For simplicity, we perform the selection in a single band, although some surveys perform the detection in a combination of bands. For instance, \citet{DES_2016} use a combination of the r+i+z bands for detection, which we approximate in the following with a simple i-band selection. 
\end{redblock}

\subsection{\label{sc:DES} Wide-like sample compared with DES Balrog}

We test the multi-passband transfer method using the DES Y3 Balrog catalogue \citep{Everett_2022}, which contains synthetic galaxies injected into single-epoch DES images and processed through the same pipeline as real wide-survey data. Out of 26 million objects, 11 million have been injected into co-add images, detected, and subsequently catalogued. The synthetic sources are created based on the DES Deep Field sources \citep{Hartley2022}, providing both high-S/N Deep photometry and realistic wide-like photometry from the reprocessed images. However, some injected sources originate from defective regions, for example, at the detector edges in the Deep field, or are re-injected into problematic areas in the Wide survey, leading to unrealistic photometric properties in either regime. To mitigate these effects, we apply quality cuts to the DES Deep catalogue, selecting only sources with \texttt{kNN\_class = 1} (galaxy classification), \texttt{badpix\_frac < 0.75} (low fraction of bad pixels), and $\log_{10}(\texttt{bdf\_T}) < 0.1\,\log_{10}(\texttt{bdf\_flux\_i})^2 + 0.2$ (shear sample selection), where \texttt{bdf\_T} is the model area and \texttt{bdf\_flux\_i} is the $i$-band flux from the Bulge+Disk model. These cleaned Deep sources are then cross-matched to the injected sample in the Balrog catalogue to eliminate objects that may affect the subsequent transformation.

To remove Deep sources that were injected into problematic Wide-field regions, we apply additional selection cuts on the Balrog catalogue. Specifically, we require \texttt{meas\_FLAGS\_GOLD\_SOF\_ONLY = 0} (no processing issues), \texttt{meas\_cm\_flags = 0} (successful composite model fit), all elements of \texttt{meas\_psf\_flags} to be zero (reliable PSF fitting), and a compact size cut on the object given by $\log_{10}(\texttt{meas\_cm\_T}) < 0.1\,\log_{10}(\texttt{flux\_i})^2 + 0.2$, where \texttt{meas\_cm\_T} is the deconvolved area and \texttt{flux\_i} is the measured $i$-band flux. We then select 500\,000 sources to form the Balrog Deep catalogue and 2\,000\,000 sources for Balrog Wide catalogue. For the Balrog Deep catalogue, we use \texttt{true\_bdf\_flux} and \texttt{true\_bdf\_flux\_err} as the fluxes and associated errors. For the Balrog Wide catalogue, we adopt \texttt{meas\_cm\_flux} and extract errors from the diagonal elements of \texttt{meas\_cm\_flux\_cov}. We apply the same selection to the DES Y3 Gold product and randomly select 2\,000\,000 sources to form the DES Wide catalogue; an additional, independent DES Wide catalogue of the same size is prepared for validation purposes. Two wide-like samples are generated from the Balrog Deep catalogue using the SPT (Sect. \ref{sc:gaussian_projection}) and MPT (Sect.~\ref{sc:multiband}) methods. Ten realisations of each transfer have been performed.

We present the flux and flux error distributions across the four passbands in Fig.~\ref{fig:compare_balrog} for the different methods compared with those of the DES Wide data set. To evaluate the performance of the transfer methods, we adopt the Wasserstein distance \citep{Kantorovich_1960}, which is a metric that quantifies the similarity between two multi-dimensional probability distributions, with smaller values of the metric indicating a closer match. In this analysis, we compare the Balrog Deep sample (before transformation), the Balrog Wide sample (generated via image simulation), and the samples transformed using both the SPT and MPT methods against the DES Wide sample. We also use the independently drawn DES Wide catalogue, which should match the original DES Wide sample by construction, to establish a benchmark for the Wasserstein distance. The Wasserstein distances are computed across three feature spaces: all flux bands, all flux error bands, and the full space combining both fluxes and flux errors. The results of this comparison are presented in Table\,\ref{tab:swd_results}.

\begin{table}[tb]
\centering
\caption{Wasserstein distance measured between the DES Wide reference subset and five comparison samples across three feature spaces: fluxes, flux errors, and the combined flux and error space. “Balrog Deep” and “Balrog Wide” correspond to Balrog simulations with deep-like and wide-like noise levels, respectively. “SPT” and “MPT” refer to the transfer methods introduced in Sect.~\ref{sc:Transferring a deep data into wide-like}. “DES Wide” denotes a second, independent subset of the DES Wide sample, used as a benchmark.}

    \begin{tabular}{lccc}
    \hline
    \hline
    \rule{0pt}{1.1em}{Sample} & {Fluxes} & {Errors} & {Fluxes $+$ Errors} \\
    \hline
    \rule{0pt}{1.1em}Balrog Deep    & 0.37 & 3.58 & 3.68 \\
    Balrog Wide    & 0.34 & 1.00 & 1.14 \\
    SPT    & 0.35 & 1.12 & 1.35 \\
    MPT  & 0.34 & 0.53 & 0.78 \\
    DES Wide    & 0.31 & 0.37 & 0.61 \\
    \hline
    \end{tabular}

\label{tab:swd_results}
\end{table}

While the DES Wide to DES Wide comparison shows, as expected, the lowest Wasserstein distance for the three feature spaces, we find that all methods perform similarly in terms of reconstructing the flux-flux correlations. This is understandable, as the correlation is mostly due to the nature of the sources, and in particular to their SEDs. However, when considering errors, we find that all transfer methods can considerably reduce the Wasserstein distance compared to the Balrog Deep case (no transformation). Nevertheless, while Balrog Wide and the SPT methods are very similar, the MPT method significantly improves over the two other methods. This is achieved in spite of the huge computational cost of the Balrog Wide transfer method.

\section{\label{sc:nz_calibration}Photometric redshift distribution calibration}
To assess the performance of the MPT method to improve the calibration of the $n(z)$ with the SOM, we implement it in the calibration process and apply it to the \Euclid Flagship Simulation \citep{EuclidSkyFlagship}. This state-of-the-art galaxy mock catalogue was developed to support \Euclid's scientific exploitation and to test data processing and calibration algorithms designed to meet the mission's scientific goals. The Flagship Simulation is based on an $N$-body simulation with four trillion particles \citep{Potter2017}, generating a light cone populated with galaxies through halo occupation distribution \citep{Berlind_2002,Zheng_2005} and abundance matching techniques \citep{Vale_2004,Conroy_2006}. The final data set contains 3.4 billion galaxies down to a magnitude of  $\HE < 26$, covering one octant of the sky up to redshift $z = 3$, with photometric measurements across multiple bands, observed (cosmological and peculiar) redshifts, and true (cosmological only) redshift values. In this test, we do not use real observational data, because ground-truth values are required to validate the performance. Therefore, using the Flagship Simulation provides us with a controlled environment in which the true bias can be measured.
    
By introducing the MPT method into the Flagship photometry, we can evaluate its effect on the accuracy of the recovered redshift distribution. Since the Flagship Simulation realistically incorporates galaxy clustering and survey selection effects, it provides a controlled and yet representative environment to test the robustness of our approach.

\subsection{Data preparation}
We select galaxies from a large sky patch defined by \( 220^\circ < \mathrm{RA} < 230^\circ \) and \( 0^\circ < \mathrm{Dec} < 10^\circ \), comprising 20 million objects from the Flagship Simulation version 2.1 Wide catalogue gathered from CosmoHub \citep{Carretero_2017, Tallada_2020}. Photometric measurements are available across the following bands: \Euclid \IE \YE, \JE, and \HE bands, and DES \( g, r, i, z \) bands. The \Euclid Flagship Simulation provides the true fluxes for each galaxy, which form the basis of our test. However, to obtain realistic observed fluxes in both Deep and Wide configurations, it is necessary to add photometric noise.
    
\subsubsection{Preparation of the test catalogues}
In the Flagship Simulation, photometric noise is added independently to each band, which does not reproduce the inter-band noise correlations present in real observations. While this approach is suitable for many validation tasks, it is not adequate for testing our method, which is designed to preserve correlations between all combinations of fluxes and flux errors across bands. To address this, we add to the true fluxes photometric noise due to the Poisson process and background fluctuations. Poisson noise depends on the count-to-flux conversion factor in each band and increases as the square root of the true fluxes. For the background noise, we define a correlation matrix of the uncertainties between the different passbands, with diagonal elements equal to 1 and off-diagonal elements arbitrarily set equal to 0.37 to introduce mild correlations in the amplitudes of the uncertainties. The value of 0.37 is motivated by empirical correlations measured in the Euclid Quick Data Release \citep{Q1-TP001}. Each uncertainty is then applied individually to their respective fluxes. We generate both Wide and Deep sample objects using the original 20 million sources, with the Wide and Deep samples assigned different Poisson and background noise levels. The reconstructed flux and flux error distributions of the new Wide sample catalogue match those of the Flagship Simulation 2.1 Wide catalogue’s observed fluxes and flux errors, except that the latter ones do not include error correlations. The Deep catalogue is created with a depth equivalent to 40 times the exposure time of the Wide survey. For validation purposes, we use the observed redshift ($z_{\mathrm{obs}}$) provided in the Flagship Simulation as the reference values for the redshift, which correspond to the true galaxy redshift, including the contribution from the peculiar velocity. We follow the same procedure as described in Sect.~\ref{sc:multiband} to transfer all deep-sample objects to wide-like objects. In the end, we construct three catalogues: the Deep catalogue, the Wide catalogue, and the wide-like catalogue obtained from the Deep catalogue, which we refer to hereafter as the ``MPT-Mock catalogue''. 

\subsubsection{Photometric redshift computation\label{sec:photoz}}
Redshift distribution calibration requires a point estimate of the redshift for tomographic-bin assignment. Since our catalogues are generated by creating new realisations of the observed fluxes from the true fluxes, the original photo-$z$ values from the Flagship Simulation cannot be used. As our goal is to evaluate the performance of the $n(z)$ calibration enabled by the MPT method, we aim to minimise the impact of redshift estimation on this validation. To keep the test focused, we adopt a simple machine-learning approach. We use a Random Forest (RF) regression \citep{Breiman_2001}, implemented using \texttt{RandomForestRegressor} from the \texttt{scikit-learn} Python package \citep{scikit-learn} with {\tt max\_depth} set to $180$ and all other parameters set to their default values, to compute the point estimates of the photo-$z$s. The regressor is trained using the seven photometric bands $griz\YE\JE\HE$ from the Deep catalogue as input features and the corresponding spectroscopic redshifts as labels. The Deep catalogue is randomly split, with 70\% of the objects used for training and the remaining 30\% for validation. The resulting training and validation samples are obviously not consistent, because their selection functions and photometric properties differ. An estimator trained on the deep catalogue therefore cannot perform as accurately when applied to the wide data.
\begin{figure}[tbp!]
    \centering
    \includegraphics[angle=0,width=1.0\hsize]{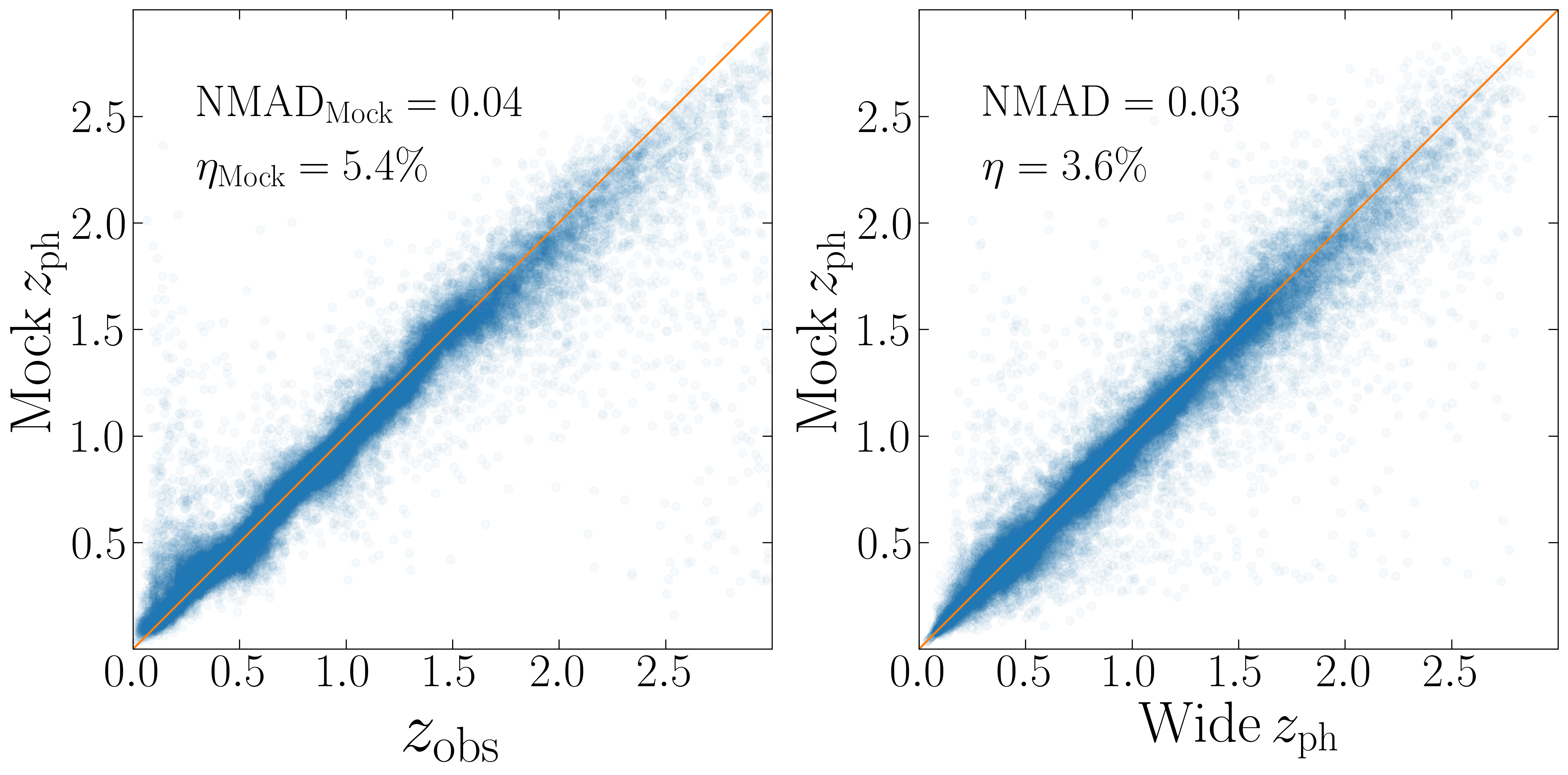}
    \caption{{Left}: Comparison between RF photo-$z$ estimates and spectroscopic redshifts for objects in the MPT-Mock sample. A magnitude cut of $\IE\!<\!25$ and an S/N$\,\geq\!10$ cut on the \IE-band photometry are applied. {Right}: Comparison of RF photo-$z$ estimates between matched objects in the Wide and MPT-Mock catalogues. The NMAD of the residuals and the outlier fractions are indicated in the figures, sources with $|z_{\mathrm{ph}}-z_{\mathrm{obs}}|>0.15(1+z_{\mathrm{obs}})$ being defined as outliers. }
    \label{fig:photozspecz}
\end{figure}

\begin{figure*}[!htb]
    \centering
    \includegraphics[angle=0,width=0.85\hsize]{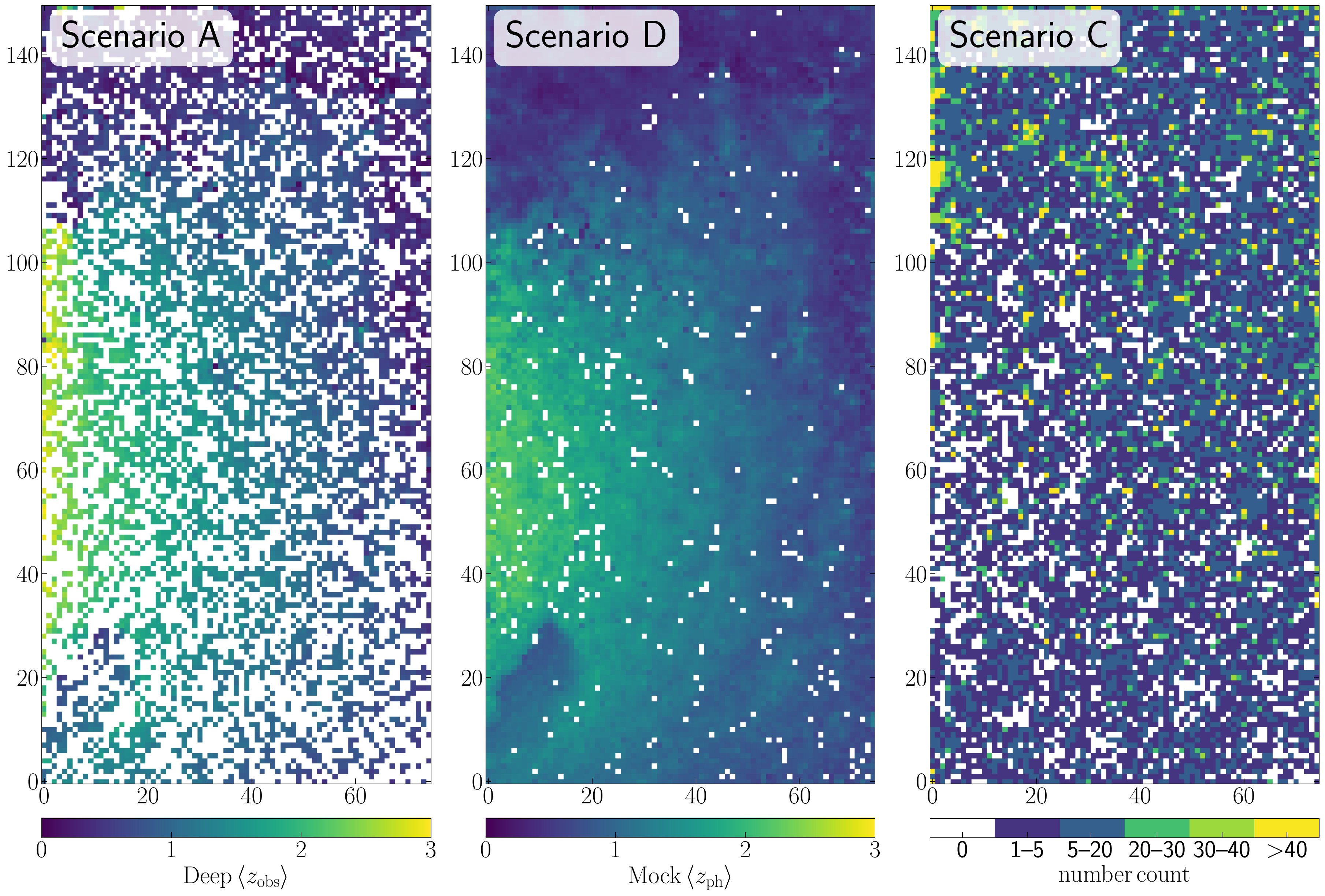}
    \caption{SOM constructed from the Deep sample populated by redshift and objects number count. The SOM is constructed using 500\,000 deep-sample objects with dimensions 75\,$\times$\,150. \emph{Left:} SOM populated with the 15\,000 deep-sample objects that have $z_{\mathrm{obs}}$ information, where each cell shows the mean $z_{\mathrm{obs}}$ of the samples it contains. This shows the original \citet{Masters_2015}'s method. \emph{Middle}: SOM populated using the 500\,000 MPT-Mock-sample objects, which are the wide-like counterparts of the deep-sample objects used to construct the SOM. Each cell shows the mean photo-$z$ of the samples it contains (Scenario D, photo-$z$ map). \emph{Right}: SOM populated with the MPT-Mock-sample objects corresponding to the 15\,000 deep-sample galaxies with $z_{\mathrm{obs}}$ information, each replicated through 50 independent realisations, in total 750\,000 objects. The colour scale indicates the number of objects per cell (Scenario C, MPT-Mock-sample occupation for the $z_{\mathrm{obs}}$-sample projection). White colour indicates the empty cells. The middle and right plots reflect the MPT-Mock-sample object onto a deep-sample constructed SOM.}
    \label{fig:dan_som}
\end{figure*}  

We assess the photo-$z$ performance by comparing the predicted photo-$z$ values to the observed redshifts $z_{\mathrm{obs}}$. Figure~\ref{fig:photozspecz} shows the comparison between the estimated photo-$z$ and the $z_{\mathrm{obs}}$ for the MPT-Mock-sample objects, with a normalised median absolute deviation (NMAD) of $0.04$ and an outlier fraction of $5.4\%$, for the samples that satisfy $\IE < 25$ and a signal-to-noise ratio $\geq 10$ in the \IE-band. However, this performance reflects both the quality of the MPT-Mock catalogue produced via MPT and the predictive power of the regression model. To disentangle the impact of the transfer method itself, we compare photo-$z$ estimates of the same sources in the MPT-Mock and Wide catalogues. The MPT-Mock catalogue contains wide-like objects generated by degrading deep photometry using our method, while the Wide catalogue comprises the same objects as observed in the wide survey. By cross-matching the two sets and comparing their respective RF photo-$z$ estimates, we find that they match very well, without evidence of additional bias, although, as expected due to the randomisation of the fluxes, there is some scatter, with $\text{NMAD}=0.03$ and an outlier fraction of 3.6\%. For the wide-sample objects, the NMAD with respect to the $z_{\mathrm{obs}}$ is $0.03$ and the outlier fraction is $3.4\%$. The corresponding figures for the MPT-Mock sample are $\text{NMAD}=0.04$ and outlier fraction is $5.4\%$. We find that the photo-$z$ performance obtained using the MPT-Mock sample shows a slight degradation compared to that of the Wide sample, when evaluated against the observed redshift $z_{\mathrm{obs}}$ as the ground truth.
The photo-$z$ performance also reflects the intrinsic uncertainty of the RF model in the Deep sample. The photo-$z$ of the Deep sample with respect to $z_{\mathrm{obs}}$ have an NMAD of $0.02$ and an outlier fraction of $0.8\%$. Adding quadratically NMAD, we find that the NMAD of the MPT-Mock sample can be well explained by the additional noise due to the deep photometry, while the outlier fraction is a bit higher than the sum of the outlier fractions in the Deep and in the Wide samples.

\subsection{\label{sc:calibration}Calibration using self-organising maps}

To introduce MPT for $n(z)$ calibration, we adopt the calibration method proposed by \cite{Masters_2015} as our baseline. This method has been further developed in subsequent applications of the SOM for the $n(z)$ calibration \citep{Wright_2020, Myles_2021, Roster_2025}. We explore below several modifications making use of our MPT-Mock catalogue to identify which ones have the most significant impact on the resulting $n(z)$ reconstruction. We also use realistic numbers of reference objects and spectroscopic redshifts to avoid achieving overoptimistic performance. 

In our implementation of the original SOM-based calibration of \cite{Masters_2015}, which we refer to as Scenario A, we randomly select 500\,000 galaxies from the Deep catalogue to form the Deep subset, of which 15\,000 galaxies are randomly selected and used as the Deep calibration subset with known $z_\mathrm{obs}$; these numbers have been chosen to reflect realistic numbers for \Euclid \citep{Masters_2017, Masters_2019, Stanford-EP14}. These subsets are used to calibrate the $n(z)$ distribution of the Wide sample, from which we randomly select 2\,000\,000 galaxies from the Wide catalogue to form the Wide subset. In all these samples, as well as in those defined below, we have applied a selection function to keep objects with the \IE\ S/N~$\ge\,10$, which is the threshold originally set to obtain a good quality shape measurement for the weak lensing analysis. We create a SOM with dimensions 75\,$\times$\,150 and train it with the Deep subset (see Appendix\,\ref{apdx:A} for technical details about our SOM). A modification that we introduce in all our scenarios is that we use the ratios of the $g$, $r$, $i$, $z$, \YE, \JE, and \HE\ fluxes to the \IE\ flux, instead of magnitude-based colours. Using these flux ratios allows the values used to train the SOM to remain roughly Gaussian distributed. 
We then project all objects with $z_{\mathrm{obs}}$ from the Deep calibration subset onto the trained SOM.
Each SOM cell containing at least one $z_{\mathrm{obs}}$ is assigned its average value in the cell, forming what we refer to as the $z_{\mathrm{obs}}$ map, and empty cells are discarded, as they are uncalibratable. We define ten tomographic bins with redshift boundaries at 0, 0.25, 0.5, 0.75, 1, 1.25, 1.5, 1.75, 2, 2.25, and 2.5. The tomographic-bin map is then constructed by assigning each SOM cell to a bin based on its mean $z_{\mathrm{obs}}$. For example, a cell with a mean $z_{\mathrm{obs}}$ of 0.1 is assigned to the first tomographic bin. We project objects from both the Deep calibration subset and the Wide subset onto the trained SOM and assign each object to a bin based on its position in the tomographic-bin map. The weight of each cell is set by the number of wide-sample objects it contains, while the $z_{\mathrm{obs}}$ values of the deep-sample objects in that cell define a normalised redshift distribution in cell $i$, that we write $P_i(z)$. For each tomographic bin, we construct a calibrated $n(z)$ of the source in the Wide sample as  

\begin{equation}\label{eq:cali_nz}
n(z) = \sum_{i=1}^{m} N_{\mathrm{wide},i}\, P_i(z)\,,
\end{equation}  

\noindent where $m$ is the total number of cells in the tomographic bin and $N_{\mathrm{wide},i}$ is the number of wide-like samples in cell $i$. In this test setup, the Wide sample is constructed from the Flagship Simulation, so that the ground truth redshift distributions are known, which can be compared with our reconstructed $n(z)$.

As noted in \citet{Roster_2025}, using $z_{\mathrm{obs}}$ to define tomographic bins can affect the calibration performance due to the double use of the $z_{\mathrm{obs}}$ information -- first when constructing the tomographic-bin map, and then again when constructing the $n(z)$ distribution of the weighted deep-sample objects. As a first modification to the original method of \citet{Masters_2015}, we define Scenario B, where the tomographic binning is based on the mean of the photo-$z$s of the 500\,000 deep-sample objects in each cell, the photo-$z$ values being computed using the method presented in Sect.~\ref{sec:photoz}.

We then implement four additional configurations of the calibration procedure based on Scenario B, where the MPT-Mock sample is used at different stages. The configurations for the following scenarios are also listed in Appendix\,\ref{apdx:B}.

\bi
    \item Scenario C: The SOM is constructed, and the tomographic-bin map is defined, as in Scenario B. The Deep calibration subset is then transformed fifty times using MPT to form the MPT-Mock calibration subset, which then consists of $50\times 15\,000$ objects. The MPT-Mock calibration subset is then projected onto the SOM, and the weights of the SOM cells are calculated according to the MPT-Mock sample's number density. The corresponding $z_{\mathrm{obs}}$ measurements from the MPT-Mock calibration subset are used to calculate the average $z_{\mathrm{obs}}$ in each cell. 
    \item Scenario D: Building up from Scenario B, the tomographic-bin map is defined by the average of the photo-$z$s of the MPT-Mock subset objects. 
    \item Scenario E: The SOM is trained using the MPT-Mock subset instead of on deep-sample objects from the Deep subset. 
    \item Scenario F: All the steps added in Scenarios C, D, and E are implemented: The Deep sample is fully replaced by the MPT-Mock sample throughout the pipeline. The SOM is trained on the MPT-Mock-sample objects from the MPT-Mock subset; the photo-$z$ map and the tomographic bins are defined using photo-$z$ values from the MPT-Mock subset, and the weight of each cell is determined from the number of MPT-Mock-sample objects from the MPT-Mock calibration subset.
    \item Scenario G: Similar to Scenario F, except that the objects are assigned to the tomographic bins based on their individual wide-sample photo-$z$, following \citet{Roster_2025}.
\ei

Figure~\ref{fig:dan_som} illustrates the redshift and occupation maps under different scenarios: a Deep SOM trained with the Deep-sample objects populated with the Deep calibration subsets shows a large fraction (46.5\%) of uncalibratable cells. In Scenario D, the same SOM is populated by MPT-Mock-sample objects for the definition of the tomographic bins, showing that the distribution of redshifts on the SOM matches that of the $z_\mathrm{obs}$. The occupation of the MPT-Mock calibration subset in Scenario C leaves a much smaller fraction of the SOM cells (6.7\%) not covered with $z_{\mathrm{obs}}$, thanks to the multiple realisations of the transfer.

\begin{figure*}[!tb]
    \centering
    \includegraphics[angle=0,width=0.85\hsize]{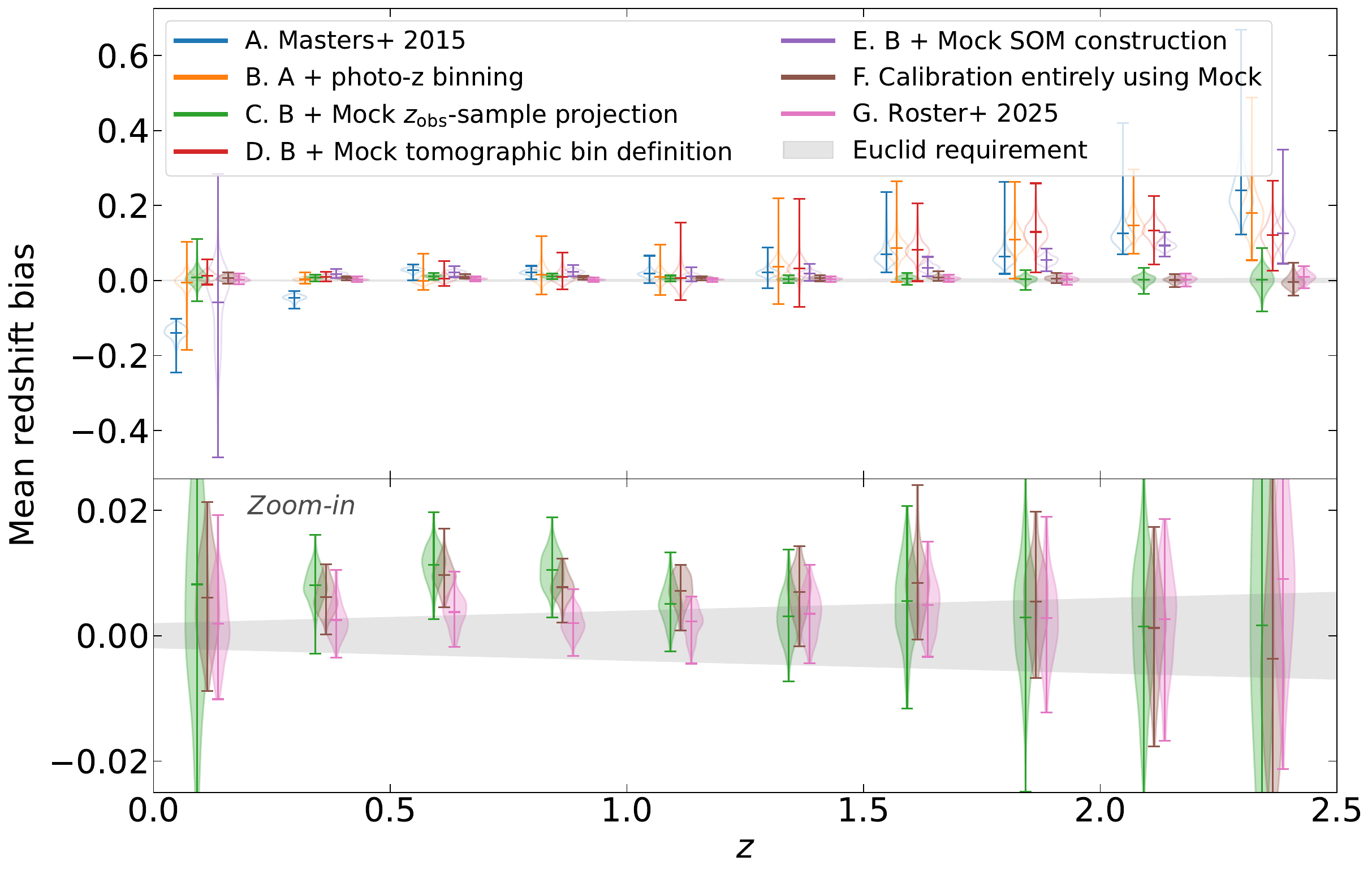}
    \caption{Mean redshift bias for different configurations of the calibration pipeline, shown as a function of redshift for ten equal-$z$ tomographic bins. The bias is computed relative to the true mean redshifts of the wide-sample $n(z)$ distributions. Violin points represent the distribution of 70 realisations, where violin points with face colour indicate calibration that used MPT-Mock-sample objects for projection and $n(z)$ reconstruction and points with no face colour indicate the projection and reconstruction are done by using deep-sample objects. The grey shaded region indicates the \Euclid requirements for $n(z)$ accuracy in weak lensing cosmology. Blue data points represent Scenario A (original \citealt{Masters_2015} method); Orange data points show Scenario B (photo-$z$s are used to define the tomographic binning); Green ones Scenario C ($z_{\mathrm{obs}}$ projection replaced with MPT-Mock-sample objects); Red data points show Scenario D (Tomographic bin defined by MPT-Mock-sample objects); Purple data points are Scenario E (SOM constructed by MPT-Mock-sample objects); Brown data points show Scenario F where full calibration is based on MPT-Mock-sample. The pink data points correspond to Scenario G, use the per-object photo-$z$ binning introduced in \citet{Roster_2025}. The data points have been slightly shifted along the $x$-axis for clarity.}
    \label{fig:nz}
\end{figure*}

Figure \ref{fig:nz} shows the bias in the mean redshift of each tomographic bin relative to the ground truth distribution for the seven calibration scenarios discussed above. To estimate the uncertainty on each measurement, we repeat the entire data selection and calibration process 500 times by drawing sources at random from the parent catalogues and compute the standard deviation and distributions of the resulting mean redshifts. Table \ref{tab:nz_result_metric} shows the average scaled biases (i.e.\ divided by $1+z$) in each bin. For the distributions that overlap with the \Euclid requirements, we compute the corresponding probability in percentage. We find that only Scenarios C, F, and G have a non-negligible probability to meet the requirements in most tomographic bins, Scenario G being the only one to have a non-negligible probability in all bins (larger than 30\%).

\begin{figure}[tbp!]
    \centering
    \includegraphics[angle=0,width=0.85\hsize]{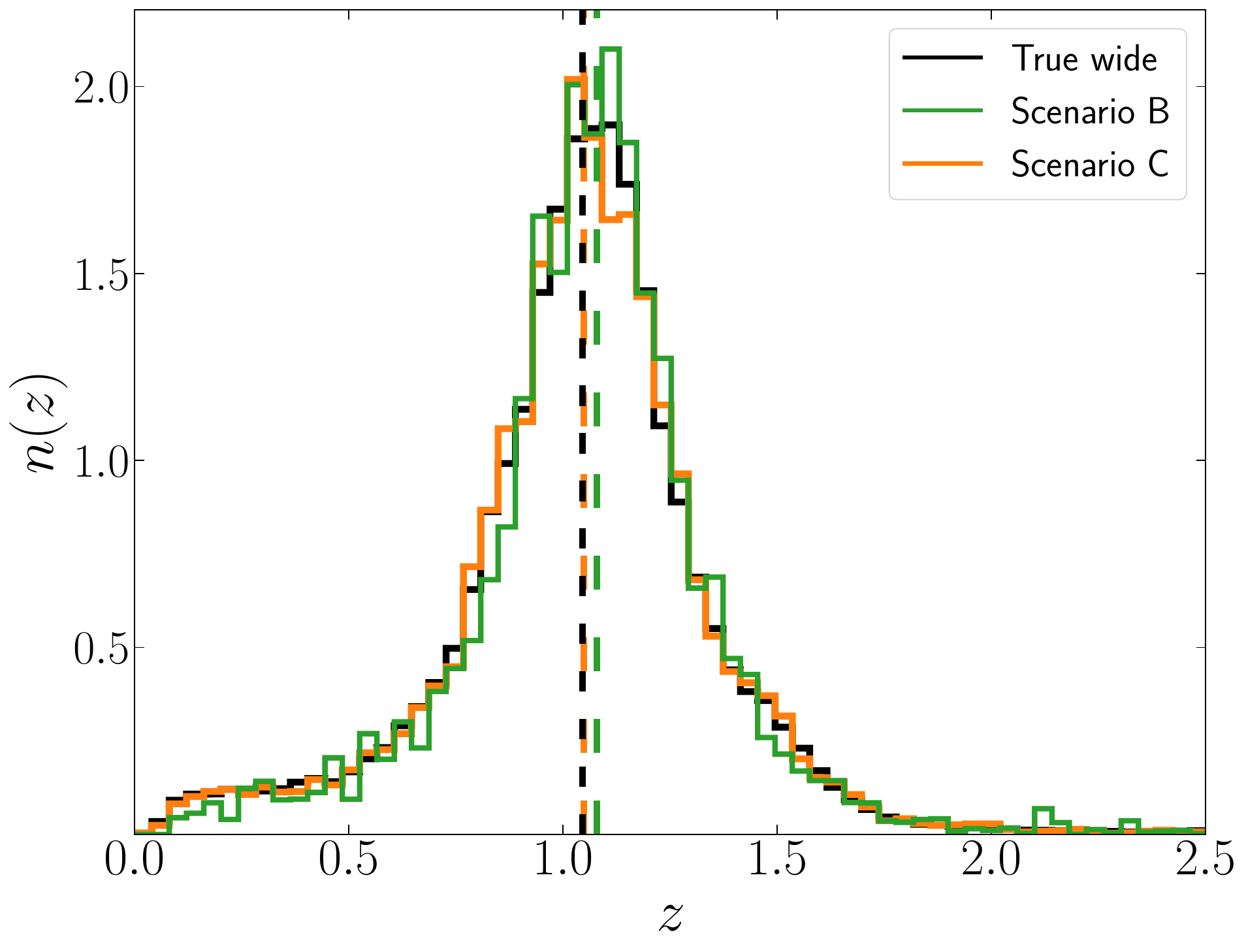}
    \caption{$n(z)$ distributions for the tomographic bin in the redshift range $1.0 < z \leq 1.25$. The black line shows the true $n(z)$ distribution of the wide-sample objects (mean $z=1.045$, variance $0.098$), with the dashed vertical line indicating its mean redshift. The orange line represents the distribution obtained when projecting $z_{\mathrm{obs}}$ from MPT-Mock-sample objects onto the SOM (Scenario C; mean $z=1.048$, variance $0.099$), where both the SOM and tomographic binning are based on the MPT-Mock sample. The green line shows the result from Scenario B (mean $z=1.079$, variance $0.113$).}
    \label{fig:nz_tomo}
\end{figure}

Figure \ref{fig:nz_tomo} shows one of the resulting $n(z)$ distributions for the tomographic bin in the redshift range $1.0 < z \leq 1.25$. The true wide-sample $n(z)$ distribution is well reproduced with the calibration method from Scenario C, while the calibration method from \citet{Masters_2015} with photo-$z$ binning (Scenario B) underestimates the width of the distribution. Quantitatively, we measure a variance of 0.098 for the true $n(z)$, 0.113 with Scenario B, and 0.099 with Scenario C. Thus, an error of 15\% is obtained with Scenario B, which reaches only 1\% with Scenario C. Scenario F uses a different SOM and a different binning scheme, so its true $n(z)$ distributions differ from those in Scenarios B and C. We find that the variance in the same bin in Scenario F is 0.089, while the true $n(z)$ has a variance of 0.088, again providing an error of about 1\% only.

\renewcommand{\arraystretch}{1.3}

\begin{table*}[t]
\centering
\caption{Fraction of the tests that lie within the \Euclid requirements. When this fraction is 0, the table provides instead the average bias divided by $1+z$, which can be directly compared to the \Euclid\ requirements of $0.002$; the value is then presented in square brackets.}
\label{tab:nz_result_metric}
\setlength{\tabcolsep}{1.5pt}

\begin{tabular}{p{0.5cm}cccccccccc}
\hline
\hline
  &
$[0,0.25)$ & $[0.25,0.5)$ & $[0.5,0.75)$ & $[0.75,1.0)$ &
$[1.0,1.25)$ & $[1.25,1.5)$ & $[1.5,1.75)$ &
$[1.75,2.0)$ & $[2.0,2.25)$ & $[2.25,2.5)$ \\
\hline
A & [0.124] & [0.034] &  [0.017] &  [0.011] &  \phantom{0}3\% &  \phantom{0}2\% &  [0.027] &  [0.022] &  [0.040] &  [0.071] \\
B & 13\% & 44\% &  36\% &  16\% &  20\% &  \phantom{0}2\% &  [0.033] &  [0.038] &  [0.047] &  [0.053] \\
C & 13\% & \phantom{0}2\% &  [0.007] &  [0.006] &  38\% & 66\% & 43\% & 48\% & 41\% & 21\% \\
D & \phantom{0}9\% & \phantom{0}3\% &  34\% &  23\% &  18\% &  \phantom{0}5\% &  [0.031] &  [0.045] &  [0.043] &  [0.036] \\
E & \phantom{0}5\% & [0.013] &  [0.014] &  [0.013] &  \phantom{0}8\% &  \phantom{0}2\% &  [0.013] &  [0.019] &  [0.030] &  [0.037] \\
F & 18\% & 10\%  & \phantom{0}3\% & \phantom{0}7\% &  20\% & 39\% & 48\% & 57\% & 51\% & 25\% \\

G 
& 36\% & 57\% &  41\% &  77\% &  76\% &  61\% &  51\% &  63\% &  57\% &  31\% \\

\hline

\end{tabular}
\end{table*}

\section{\label{sc:discussion} Discussion}

\subsection{Performance of the multi-passband transfer}
Because of the differences in the survey strategies, there is a mismatch in photometric properties between deep- and wide-sample objects. A common approach to mitigate this mismatch is to degrade deep-sample objects by adding random scatter to approximate the noise properties of Wide observations. However, this simple noise-adding method disrupts the correlation structure imprinted in the measurements, particularly the correlations between flux errors across different passbands. 

In this work, we introduce and test successfully a multi-passband transfer method designed to simulate wide-like photometry from Deep observations. This catalogue-level transfer serves as a practical alternative to image-based simulations such as Balrog for matching deep- and wide-sample objects, offering a computationally efficient approach capable of processing one million sources in eight bands in five minutes with an Intel(R) Xeon(R) Gold 5218 CPU. 

In addition, with the metrics we used here, we find that MPT recovers the statistical properties of the wide sample slightly better than Balrog. This arises possibly because Balrog injects sources using a synthetic model, and their photometry is extracted with a similar model, making the injected sources less realistic than actual wide-sample objects. Consequently, the flux errors measured for Balrog Wide objects are smaller than they would be if the same objects appeared in real DES Wide observations. This discrepancy can degrade the performance of Balrog Wide in recovering the wide-sample error distributions.

With the eight photometric bands of the Flagship Simulation, the transfer is performed in 16 dimensions. In principle, our probabilistic sampling should include the entire wide-sample data set, without pre-selection of the neighbours. However, since wide surveys typically contain millions to billions of sources, this approach becomes computationally infeasible. We use a k-d tree to identify the nearest neighbours of each deep-sample object, and then perform a likelihood-based sampling within this local subset. The use of nearest neighbours as a pre-selection strategy significantly improves the efficiency with a moderate loss of precision. We find that setting $k=50$ provides a good balance between transfer accuracy and computational performance.

Figure \ref{fig:compare_balrog} shows that MPT successfully reproduces the flux and flux-error distributions of wide-sample objects by degrading deep-sample objects. While matching the distributions in flux and uncertainty space is a necessary first step, the true value of this method lies in its ability to support downstream scientific applications. One such application is photometric redshift estimation. The right panel of Fig.~\ref{fig:photozspecz} illustrates the implementation of this transfer function. The photo-$z$ values estimated from MPT-Mock-sample objects closely match those obtained from the same objects observed directly in the Wide data set. The NMAD (0.03) and outlier fraction (3.4\%) for the wide-sample objects indicate the upper limit that can be achieved by degrading the corresponding deep-sample objects. The MPT-Mock-sample object shows a slightly worse performance than with NMAD equal to 0.04 and outlier fraction equal to 5.4\%, but this can be reasonably well explained by the intrinsic scatter in the photo-$z$ of the Deep sample objects. In addition, we do not find any evidence of bias in the photo-$z$ introduced by the MPT. This shows that the transformation does not significantly degrade information compared to wide-sample objects, and enables a reliable estimation of derived properties, such as redshift, highlighting its effectiveness beyond simple distribution matching.

\subsection{Impact on the redshift distribution calibration}
A basic requirement for an accurate photometric redshift calibration is the quite intuitive idea of \citet{Lima_2008} that the colour distributions of test and reference objects should be identical. Since this is an extremely hard requirement, the SOM is used to partition the colour space into cells where occupation of both samples can be scaled. The properties of the photometric uncertainties are also a significant part of the matching of the colour spaces. We address this here using MPT, which degrades the photometry of deep-sample objects, in our case from the EAFs, to the performance of the wide survey, the EWS. This mimics the situation in KiDS \citep{de_Jong_2013}, where extensive spectroscopic coverage directly overlaps the survey area, which does not exist for \Euclid. We test different ways to make use of this transfer method.

Figure \ref{fig:dan_som} shows the Deep $z_{\mathrm{obs}}$ map exhibits a large fraction of unpopulated cells, with 46.5\% of SOM cells lacking coverage, and therefore having to be discarded. This shows that the number of $z_{\mathrm{obs}}$ is insufficient to cover the colour space of wide-survey galaxies adequately. Some of these unoccupied cells might be the result of the spread in colour space due to the wide-photometry uncertainties. In Scenario C, the colour space is populated more uniformly on the SOM through the use of multiple wide-like realisations, reducing the fraction of unpopulated cells to 3.4\%. Scenarios F and G follow the same strategy and exhibit a similar improvement in SOM coverage.

Figure~\ref{fig:nz} clearly shows that MPT applied on the Deep-sample objects with $z_{\mathrm{obs}}$ turns out to be the most important step towards improving the SOM-based calibration, as MPT improves the matching of both samples in colour space. With Scenario C, on average, 27.2\% of the bins are well calibrated, and 27.8\% for Scenario F, while this value reaches 55\% for Scenario G. As argued in \citet{Roster_2025}, assigning objects to tomographic bins based on their individual photo-$z$ prevents the occurrence of situations where a SOM cell is assigned to a non-representative tomographic bin, because the taking of the mean of a bimodal distribution. The performance will improve with a larger sample of $z_{\mathrm{obs}}$ and better photo-$z$s.

Scenario F is only marginally better than Scenario C. However, conceptually, Scenario F makes more sense, especially the construction of the SOM using the MPT-mocked sample. Indeed, because of the uncertainties, the Wide-sample colour space is larger than the Deep-sample one. When the SOM is constructed using deep photometry, some wide-sample objects may end up outside of the colour space specified by the SOM, and then be projected on a random cell, thereby contaminating the $N_{\mathrm{wide},i}$ occupations. 

\begin{figure}[tbp!]
    \centering
    \includegraphics[angle=0,width=0.85\hsize]{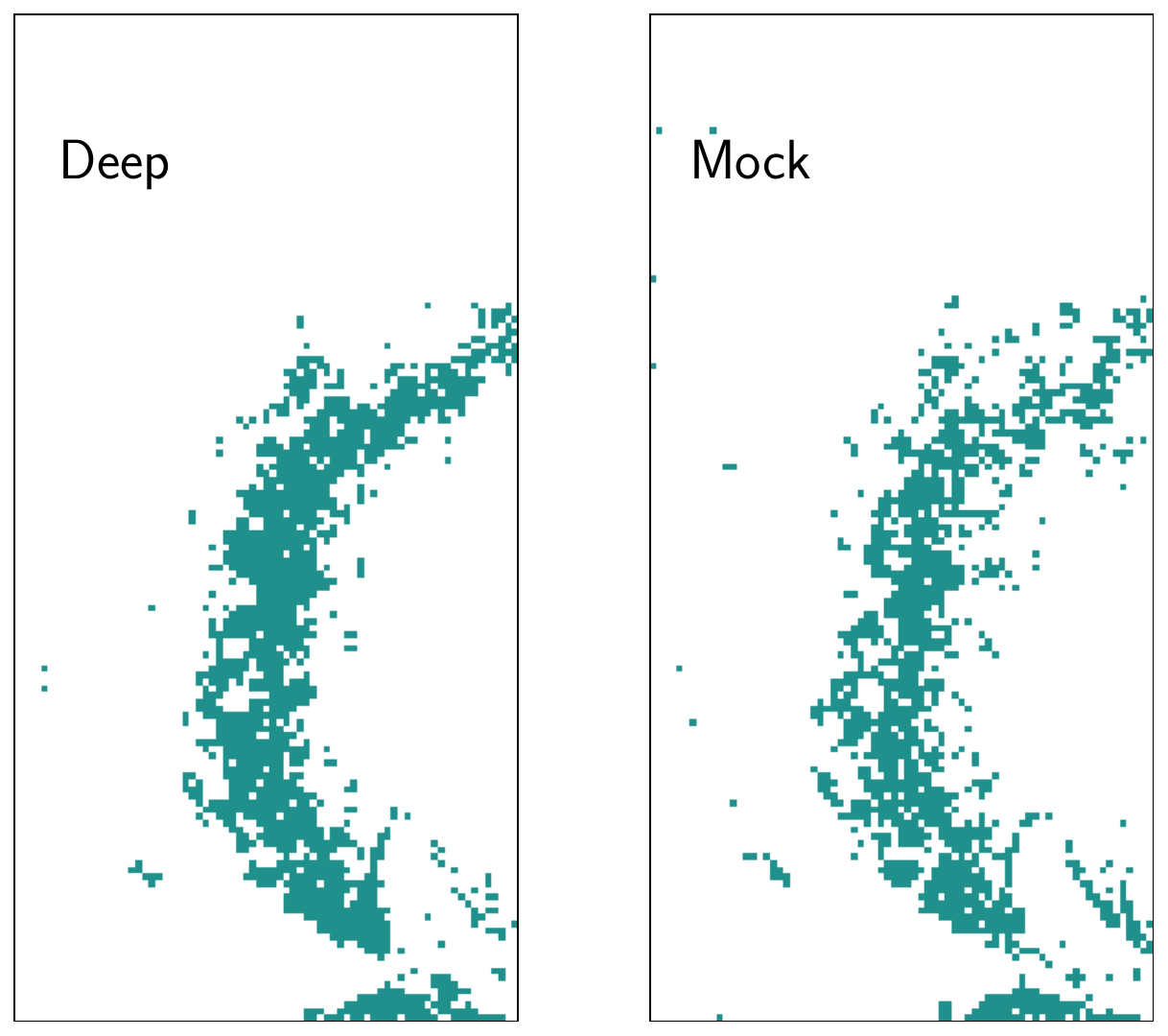}
    \caption{Comparison of tomographic-bin areas defined using deep-sample objects and MPT-Mock-sample objects. The maps show the SOM regions where the mean photo-$z$ of the cells falls within $1.25 < z \leq 1.5$, populated by deep-sample objects (left) and by MPT-Mock-sample objects (right). The MPT-Mock-sample set contains the same number of objects.}
    \label{fig:tomo_compare}
\end{figure}

With Scenario D, we test whether defining the tomographic binning using MPT-Mock-sample objects leads to any improvement. Figure~\ref{fig:tomo_compare} shows an example of the tomographic bin comparison between using the deep subset (Scenario B) and the MPT-Mock subset (Scenario D) to define the bin areas. On large scales, the resulting distributions are broadly consistent with each other. As shown in Fig.~\ref{fig:nz} and Table~\ref{tab:nz_result_metric}, the performance in this case is comparable to Scenario B, indicating that using MPT-Mock-sample objects for tomographic binning does not provide a huge improvement. This is because, while the MPT-Mock photo-$z$s are less precise, the averaging over many objects makes the cell assignments nearly identical to those obtained with deep-sample objects. This might not be true if deep and wide photometries are affected by different biases. We also notice Scenario D has a better performance at lower redshift bins. This is due to the fact that the variance of the probability distributions spans a wider range, even though the mean redshift biases further deviate from the zero offset, those data points within the \Euclid requirement will take into account the probability.

To explore the impact of the type of data used to create the SOM, we test the SOM trained on deep-sample objects (Scenario B) and the SOM trained on MPT-Mock-sample objects (Scenario E), all other steps being as in Scenario B. The photometry of deep-sample objects is close to the true values, so the SOM approximately represents the real colour space. In contrast, MPT-Mock-sample objects occupy a broader colour space due to the photometric degradation process. This corresponds to Scenario E, which actually performs the worst in our tests. This can be explained by the fact that a given location in the true colour space gets spread over many cells, while the reference objects are not, unless the technique of Scenario C is used at the same time. Scenario E creates, therefore, a mismatch between the colour space of the SOM and of the reference objects. Nevertheless, a SOM constructed from deep photometry spans a smaller region of the colour-space manifold than one built from wide photometry due to the larger uncertainties. Projecting wide-sample objects that lie outside the deep colour-space manifold onto a deep-based SOM can introduce bias. Although we do not observe a significant reduction in the $n(z)$ bias in our tests, employing a wide-like SOM may prove to be important in principle.

In summary, we confirm that matching the photometric properties of the Deep sample and the Wide sample is necessary. This is particularly true for the spectroscopic sample, as the only scenarios that produce reasonably calibrated bins are those where the Deep calibration subset is mocked using MPT and projected onto the SOM. Multiple realisations of the mocking are needed to mitigate the shot noise introduced by the small size of the sample. Even though the steps in Scenario D and E do not show very significant improvements, they can, in principle, improve the matching of the Deep and Wide samples.

\subsection{Comparison with \citet{Roster_2025}}
\label{sec:roster}
In \citet{Roster_2025}, the authors tested several modifications to the original SOM $n(z)$ calibration of \citet{Masters_2015} using the Flagship Simulations. Namely, they used a SOM constructed using wide-like photometry, and the reference objects are projected onto this SOM after additional uncertainties typical of the wide survey are applied. Each object is assigned to a tomographic bin based on its photo-$z$. Scenario G implements the different steps of the method presented in \citet{Roster_2025}, with several differences. We also focused on building a test case of the method that is as representative as possible of the \Euclid case, in terms of the number of objects in the Deep calibration sample, their distribution on the SOM, and the properties of the different photometric samples, in particular with the introduction of the correlations between errors. In \citet{Roster_2025}, wide-like uncertainties of the Deep sample are known from the simulated catalogues, which requires that the same sources are observed at both deep and wide depths. Instead, MPT allows us to construct a wide-like sample from a deep survey by matching with the properties of any wide sample, even if there is no overlap between the deep and wide samples. We also evaluate each step of the process one by one to understand its importance in the calibration.

In this paper, we have been able to achieve similar performances to those obtained in \citet{Roster_2025} in conditions that apply to the Euclid weak-lensing survey, while confirming that the matching of the photometric properties of the Deep and Wide samples is a key requirement for the SOM calibration to perform correctly. We confirm their key findings, in particular regarding the need to assign sources to tomographic bins based on their individual photo-$z$s. \citet{Roster_2025} propagated the uncertainties and biases in the tomographic-bin redshift distributions to the cosmological-parameter inference. As our results are similar, the same conclusions apply, namely that the bias on the parameter should be limited to about $0.3\sigma$ in somewhat realistic conditions.

\subsection{Implications beyond weak lensing tomography}
For weak lensing tomography, an accurate estimate of the mean redshift of the galaxy distribution is crucial to minimise systematic biases in cosmological parameter estimation. While the mean of $n(z)$ captures the primary requirement, knowledge of the full distribution remains necessary \citep{Ma_2008, Bordoloi_2010, EuclidSkyOverview}. As illustrated in Fig.~\ref{fig:nz_tomo}, MPT not only reproduces the mean redshift of the wide-sample distribution with high accuracy, but also preserves the overall shape of the $n(z)$ distribution; in particular, the second moment is recovered to within about 1\% with Scenario C or Scenario F. This demonstrates that the calibration procedure with MPT faithfully retains the statistical structure of the redshift distribution beyond just its central value. Consequently, the method has potential applications beyond weak lensing and will in particular benefit other cosmological analyses that rely on accurate modelling of the full $n(z)$ distribution, such as angular clustering \citep{Peebles_1980}.

\section{\label{sc:Conclusion} Conclusions}

In this work, we present and evaluate a multi-passband transfer method that creates wide-like photometry from deep-field observations. The method is computationally efficient and preserves the intrinsic correlations between fluxes and errors across all passbands, offering a significant improvement over naive noising techniques, and even, with the metrics we used here, over computationally heavy image-based source injection methods such as Balrog. Applied to the photometric redshift calibration method from \citet{Masters_2015} in a somewhat realistic set-up, especially in terms of the number of sources in the different samples and the presence of correlated errors, MPT is instrumental in reproducing with good performance not only the mean of the $n(z)$ redshift distribution, but also its overall shape. We find that the key role of MPT lies in the projection of the Deep $z_{\mathrm{obs}}$ sample, which allows the distribution of these objects on the SOM to match that of identical objects from the wide sample. When keeping all other steps based on deep-sample objects, this step results in the largest improvement in calibration performance. We confirm also the need to assign objects to tomographic bins based on their individual photo-$z$s. In addition, the MPT method integrates naturally into the \Euclid\ calibration pipeline, where wide-field objects lack corresponding deep-field observations. Moreover, by generating multiple realisations, this method distributes the $z_{\mathrm{obs}}$ sources around their true locations, thereby reducing the number of sparsely populated cells in the SOM. Overall, this work provides a practical solution for improving photometric redshift calibration, with broad relevance for any cosmological studies that rely on accurate measurements of the $n(z)$ distribution. While improvements are still needed after \Euclid DR1, if only by improving the quality of the reference sample, this approach offers a robust framework for future data releases.

\begin{acknowledgements}
  
\AckEC,  \AckCosmoHub
\end{acknowledgements}

%
%

\bibliography{paper_ref, Euclid}

%
\begin{appendix}
\onecolumn

\section{Technical note on the construction of the SOM\label{apdx:A}}

Another important consideration is the computational performance and stability of the SOM algorithm, especially when projecting large numbers of sample objects onto the map. Some SOM implementations in interpreted languages are too slow to deal with the millions of objects found in modern surveys. 
Some packages produce unwanted features in the trained maps, such as patterns of empty cells at regular intervals, or cells that are populated with an unrealistically large fraction of the objects. In the SOM redshift map, there are sharp discontinuities between regions of high- and low-redshift cells, leaving valleys of empty cells at the transition boundaries. These issues can severely limit the applicability of such tools in large-scale survey contexts, where both efficiency and robustness are essential. Given that the available packages do not comply with all the needs of \Euclid in terms of parallelisation, robustness and control of the input parameters, for this work, we adopted the SOM package developed by the Swiss \Euclid\ Science Data Centre, available at \url{https://github.com/astrorama/PHZ_SOMBiasCorrection/tree/0.18.1}, which has been implemented from scratch in C++ and integrated into the official \Euclid\ data processing pipeline. This implementation not only provides stable and reliable map construction, free from the aforementioned issues, but also offers fast projection speed, making it well-suited for large surveys.

During SOM construction, as in \citet{Wright_2020} and \citet{Roster_2025}, we test various input configurations, including a combination of colours plus \Euclid\ \IE\ magnitude, or more general combinations of colours plus a single passband magnitude. However, the additional dimension led to an increased number of empty cells. Moreover, due to the uncertainties, some objects contain negative flux values that cannot be converted into magnitudes or colours and have to be discarded from the SOM, which biases the colour space. Even when the flux is positive, fluxes close to the detection limit generate magnitudes that are strongly biased towards high values. However, such cuts can introduce selection biases and reduce the sample size. In contrast, using flux ratios at the SOM construction level while making sure that the flux in the denominator is well above the detection limit allows all objects, including those with negative fluxes in the other bands, to be retained. Taking into account all these considerations, we concluded that using flux ratios was more robust than using colours and magnitudes. We point out in addition that the cut on the \Euclid\ \IE\ S/N is similar to that which is applied by the \Euclid weak lensing pipeline.

\section{Calibration configurations\label{apdx:B}}
\begin{center}
\label{tab:scenario}
\begin{tabular}{c c c c c}
\hline\hline
\rule{0pt}{1.1em}Scenario & SOM construction & Binning assignment & Binning redshift & Projection sample \\
\hline
\rule{0pt}{1.1em}A & Deep       & by SOM cell     & Deep spec-$z$        & Deep       \\
B & Deep       & by SOM cell     & Deep photo-$z$       & Deep       \\
C & Deep       & by SOM cell     & Deep photo-$z$       & MPT-Mock   \\
D & Deep       & by SOM cell     & MPT-Mock photo-$z$   & Deep       \\
E & MPT-Mock   & by SOM cell     & Deep photo-$z$       & Deep       \\
F & MPT-Mock   & by SOM cell     & MPT-Mock photo-$z$   & MPT-Mock   \\
G & MPT-Mock       & by object   & photo-$z$       & MPT-Mock       \\
\hline
\end{tabular}
\end{center}

\end{appendix}
\label{LastPage}
\end{document}